\documentclass[lettersize,journal]{IEEEtran}
\usepackage{amsmath,amsfonts}
\usepackage{algorithmic}
\usepackage{algorithm}
\usepackage{array}
\usepackage[caption=false,font=normalsize,labelfont=sf,textfont=sf]{subfig}
\usepackage{textcomp}
\usepackage{stfloats}
\usepackage{url}
\usepackage{verbatim}
\usepackage{graphicx}
\usepackage{cite}
\usepackage{amssymb}
\usepackage{graphicx}
\usepackage{textcomp}

\usepackage{tikz}
\usetikzlibrary{shapes,arrows,positioning,calc}
\tikzset{
block/.style = {draw, fill=white, rectangle, minimum height=2.5em, minimum width=3em},
tmp/.style = {coordinate},
sum/.style= {draw, fill=white, circle, node distance=1cm},
input/.style = {coordinate},
output/.style= {coordinate},
pinstyle/.style = {pin edge={to-,thin,black}}}

\begin{document}

\title{Hierarchical Fuel-Cell Airpath Control: an Efficiency-Aware MIMO Control Approach Combined with a Novel Constraint-Enforcing Reference Governor}
\author{Eli Bacher-Chong\textsuperscript{1}\thanks{\textsuperscript{1}Department of Mechanical Engineering at the University of Vermont, Burlington, VT 05405, USA (e-mail: Eli.Bacher-Chong@uvm.edu)}, Mostafa Ali Ayubirad\textsuperscript{2}\thanks{\textsuperscript{2}Department of Electrical and Biomedical Engineering at the University of
Vermont, Burlington, VT 05405, USA (e-mails: Mostafa-Ali.Ayubirad@uvm.edu, Hamid.Ossareh@uvm.edu)}, Zeng Qiu\textsuperscript{3}, Hao Wang\textsuperscript{3}, Alireza Goshtasbi\textsuperscript{3}\thanks{\textsuperscript{3}Research \& Advanced Engineering at
Ford Motor Company, Dearborn, MI 48124, USA (e-mails: cqiu1@ford.com, hwang210@ford.com, goshtasb@umich.edu)}, and Hamid R. Ossareh\textsuperscript{2}
\thanks{
This research received funding from Ford Motor Company through a University Research Project grant.

© 2023 IEEE.  Personal use of this material is permitted. Permission from IEEE must be obtained for all other uses.
DOI: 10.1109/TCST.2023.3329909}
}
\maketitle

\begin{abstract}
This paper presents a hierarchical mutlivariable control and constraint management approach for an air supply system for a proton exchange membrane fuel cell (PEMFC) system. The control objectives are to track desired compressor mass airflow and cathode inlet pressure, maintain a minimum oxygen excess ratio (OER), and run the system at maximum net efficiency. A multi-input multi-output (MIMO) internal model controller (IMC) is designed and simulated to track flow and pressure set-points, which showed high performance despite strongly coupled plant dynamics. A new set-point map is generated to compute the most efficient cathode inlet pressure from the stack current load. To enforce OER constraints, a novel reference governor (RG) with the ability to govern multiple references (the cascade RG) and the ability to speed up as well as slow down a reference signal (the cross-section RG) is developed and tested. Compared with a single-input single-output (SISO) air-flow control approach, the proposed MIMO control approach shows up to 7.36 percent lower hydrogen fuel consumption. Compared to a traditional load governor, the novel cascaded cross-section RG (CC-RG) shows up to 3.68 percent less mean absolute percent error (MAPE) on net power tracking and greatly improved worst-case OER on realistic drive-cycle simulations. Two fuel cell system (FCS) models were used for development and validation, a nonlinear open-source model and a proprietary Ford high-fidelity model.
\end{abstract}


\section{Introduction}
Fuel cells (FCs) are electrochemical devices that convert chemical energy in fuel directly into electrical energy. Of the many types of fuel cells, the proton exchange membrane fuel cell (PEMFC) fueled by hydrogen shows great promise for automotive applications. This type of fuel cell generates electricity with the electrochemical reaction between hydrogen and oxygen and generates water and waste heat as byproducts. The PEMFC has attracted considerable attention because a hydrogen PEMFC can have high efficiency (50-55\%) when compared to gasoline (15-25\%) and diesel (30-35\%) internal combustion engines, and zero local emission of harmful byproducts. It also promises higher power and energy density when compared to rechargeable batteries, and fast refueling time \cite{sharaf_overview_2014}. But in practice, fuel cells require fast and robust control systems to run efficiently and reliably, so research into control and constraint management of fuel cell systems (FCS) is ongoing.

{An FCS consists of the fuel cell stack itself, as well as its auxiliary components, such as the oxygen supply system, hydrogen supply system, humidification subsystem, electrical subsystem, and thermal management subsystem}. In this paper the oxygen supply system will be referred to as the air-supply or air-path system, as oxygen in this configuration is supplied by pressurizing the ambient air through an electric compressor \cite{pukrushpan_control_2004}. As such, the purpose of the controller for the air-path subsystem is to ensure high net efficiency of the FCS and fast power output response, and to avoid compressor surge and oxygen starvation.

The FCS air-path subsystem is of particular importance for several reasons. First, the air-path system can consume about 20\% of the total FC power output via the electric compressor \cite{laghrouche_load_2013}, making it a major parasitic load. Second, the air-path dynamics are slower than many of the other subsystems, making it a bottleneck in the speed of FCS power delivery.
Third, air-path operations are subject to critical constraints such as the ones on compressor surge, oxygen excess ratio (OER), and membrane water management \cite{goshtasbi_degradation-conscious_2020}. In particular, oxygen starvation, as determined by a drop in OER, can lead to a drop in the stack voltage, therefore degrading the power output, {and in  extreme cases, this voltage drop can lead to a short circuit, potentially causing localized hot-spots to form on the surface of the FC membrane-electrode assembly (MEA).}

Various control strategies have been proposed to improve the air-path dynamics with the compressor motor as the only control input. In \cite{pukrushpan_control_2004-1}, a ninth-order nonlinear model was developed and combined with a linear quadratic Gaussian (LQG) controller to prevent oxygen starvation and track the desired net power output. A linear-quadratic-regulator/Gaussian (LQR/LQG) type controller was used and experimentally validated in \cite{niknezhadi_design_2011}. To facilitate model-based control strategies, the ninth-order model in \cite{pukrushpan_control_2004} was reduced to a simplified fourth-order model in \cite{suh_modeling_2006} and a third-order model in \cite{Reine_J_Talj_2010}. Nonlinear and adaptive control strategies using low-order models were applied to the OER problem in \cite{m_adaptive_nodate,ma_oxygen_2020,han_oxygen_2019,sankar_nonlinear_2018}.

The air-path subsystem may also contain a throttle valve downstream of the cathode to raise the upstream pressure \cite{sun_coordination_2020,chen_modeling_2014}. Higher cathode pressure can improve the stack voltage by increasing reactant concentration on the electrode surfaces \cite{goshtasbi_degradation-conscious_2020}. Pressure control has also been studied to improve compressor efficiency. In \cite{sun_coordination_2020}, a multi-input multi-output (MIMO) internal model controller (IMC) was proposed to simultaneously control flow and pressure using both the compressor and the valve. However, \cite{sun_coordination_2020} simulated their controller on an identified first-order linear model rather than a nonlinear physics-based model. Also, \cite{sun_coordination_2020} tracked the optimal operating point of the compressor, which may neglect working regions where the improvement in stack power outweighs the increase in compressor parasitic loss. In \cite{tirnovan_efficiency_2012} a feed-forward map was developed to maximize system net efficiency via selection of optimal compressor speed and valve setting, but this map is not as suitable for shaping closed-loop dynamics.

Control of the compressor motor alone is not sufficient to prevent oxygen starvation, as shown in \cite{vahidi_current_2006}. To maintain a minimum OER at all times, various strategies have been studied to manipulate the load current drawn from the fuel cell, such as model predictive control (MPC) and reference governors (RG). In \cite{vahidi_constraint_2005}, an MPC and a fast reference governor (FRG) were both designed to prevent oxygen starvation. A robust nonlinear RG was applied to the oxygen starvation problem in \cite{jing_sun_load_2005}, but this was computationally demanding due to online nonlinear simulations. In \cite{vahidi_constraint_2007}, a linear-model RG was modified with a disturbance term and applied to a nonlinear FCS model. Nonlinear RGs are computationally demanding, while more computationally efficient existing linear RGs suffer from plant-model mismatch.

The main objective of this research is to investigate a constraint-aware and efficiency-aware MIMO air-path controller to overcome the limitations of the existing literature mentioned above. As such, the original contributions of this paper are as follows:

\begin{enumerate}
    \item \textbf{FCS Analysis:} {
    We show that the dynamics of the compressor airflow and the cathode inlet pressure of the FCS air-path system are highly-coupled but not highly nonlinear, justifying a non-scheduled linear MIMO controller ({Section} \ref{section:model}).}
    \item \textbf{Efficiency-Maximizing Set-Point Map:} {We include the} cathode inlet pressure as an added degree of freedom, { and show that an optimal air-path set-point map can be generated that improves the net efficiency of the FCS.}
    (Section \ref{section:set-point-maps}).
    \item \textbf{MIMO IMC Air-Path Controller:}
    {We show that a fast tracking and} easy-to-tune MIMO IMC architecture { can be designed  using a third-order physics-based model to robustly and effectively control the FC air-path} ({Section} \ref{section:IMC}).
    \item \textbf{OER Constraint Reformulation:} A new linear formulation of the previously nonlinear OER constraint is presented to reduce mismatch between the prediction by the RG and the actual nonlinear output ({Section} \ref{section:RG-OER}). This is tested on a stack current load governor ({Section} \ref{section:RGi}).
    \item \textbf{{CC-}RG:} A novel RG algorithm, the cascaded cross-section reference governor (CC-RG) ({Section} \ref{section:RGap}) is developed for OER, which demonstrates faster net power tracking than the RG governing stack current alone, while still maintaining the OER constraint. This involves two new frameworks: \begin{itemize}
        \item \textbf{Cascade RG:} The cascade RG uses multiple {single-output} RGs in series to govern multiple setpoints, { where the order of the RGs is chosen to prioritize the control objectives }({Section} \ref{section:RG-cascade}).
        \item \textbf{Cross-Section RG:} The cross-section RG allows some of the setpoints to {slow down or speed up} (instead of only {slow down} as is the case {for a standard load governor}) through a novel approach ({Section} \ref{section:RG-CS}). { As we show, this is an important feature that minimizes the loss of power in transients due to the OER constraint.}
    \end{itemize}
\end{enumerate}
{ The entire MIMO control scheme and the {CC-}RG are tested on a ninth-order nonlinear model with stack current steps and realistic drive-cycles ({Section} \ref{section:drive-cycle}) as well as on a high-fidelity FCS model from Ford Motor Company ({Section} \ref{section:highfidelity}).}
Earlier work by the authors in the conference version \cite{bacher-chong-2022} presents the development of contributions 1 through 4 in the list above, and tests these on drive-cycle simulations using a ninth-order nonlinear plant model. This paper extends \cite{bacher-chong-2022} with contributions 5, 
{which introduces two new types of RG, presents their features and limitations, and applies them to the FCS OER problem.}

\section{Fuel Cell System Modeling and Analysis}
\label{section:model}

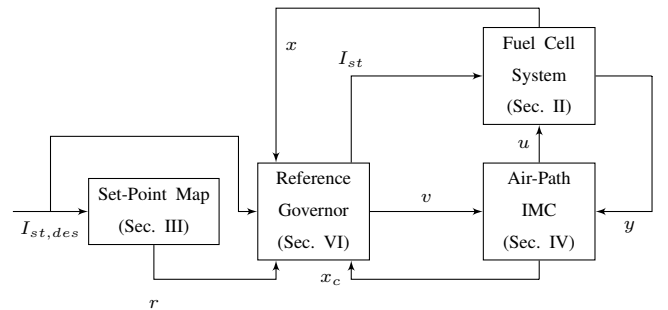
\begin{figure}
    \centering
    \vspace{0.1cm}
    \begin{tikzpicture}[auto, node distance=1.5cm,>=latex',scale=0.60]
\node [input, name=rinput] (rinput) {};
\node [block, right=1.00cm of rinput,text width=1.5cm,align=center] (sFF) {{\scriptsize Set-Point Map (Sec. \ref{section:set-point-maps})}};
\node [block, right=0.50cm of sFF,node distance=2cm,text width=1.25cm,align=center] (RG)
{{\scriptsize Reference Governor (Sec. \ref{section:RGap})}};
\node [block, right=1.50cm of RG,text width=1.25cm,align=center] (IMC) {{\scriptsize Air-Path IMC (Sec. \ref{section:IMC})}};
\node [block, above=0.50cm of IMC,text width=1.25cm,align=center] (FCS) {{\scriptsize Fuel Cell System (Sec. \ref{section:model})}};
\node [tmp, right=0.75cm of FCS] (tmp1){};
\node [tmp, right=0.25cm of RG] (tmp0){};
\node [tmp, above=0.25cm of FCS] (tmp2){};
\node [tmp, below=0.25cm of IMC] (tmp3){};
\draw [->] (rinput) -- node[yshift=-0.50cm]{\scriptsize $I_{st,des}$} (sFF);
\draw [->] (RG) -- node [name=r] {\scriptsize $v$}(IMC);
\draw [->] (IMC) -- node [name=u] {\scriptsize $u$}(FCS);
\draw [->] (FCS) -| (tmp1) |- node [name=y,xshift=-3mm] {{\scriptsize $y$}} (IMC);
\node [tmp, above=0 of RG, xshift=0.5cm] (tmpRGi) {};
\node [tmp, above=0 of RG, xshift=-0.5cm] (tmpRGx) {};
\node [tmp, below=0 of RG, xshift=0.5cm] (tmpRGxc) {};
\node [tmp, below=0.25cm of RG, xshift=-0.5cm] (tmp4){};
\node [tmp, below=0 of RG, xshift=-0.5cm] (tmpRGides) {};
\draw [->] (tmpRGi) |- node [name=w] {\scriptsize $I_{st}$} (FCS);
\draw [->] (FCS) |- (tmp2) -| node [name=x,yshift=-5mm] {\scriptsize $x$} (tmpRGx);
\draw [->] (IMC) |- (tmp3) -| node [name=xc] {\scriptsize $x_c$} (tmpRGxc);
\draw[->] (sFF) |- node[yshift=-0.50cm]{\scriptsize $r$} (tmp4) -| (tmpRGides);
\node [tmp, left=0.25cm of RG] (tmp5) {};
\node [tmp, right=0.50cm of rinput] (tmp6) {};
\node [tmp, above=1.00cm of tmp6] (tmp7) {};
\draw[->] (tmp6) |- (tmp7) -| (tmp5) -- (RG);
\end{tikzpicture}
    \vspace{-0.2cm}
    \caption{Block diagram of the proposed fuel cell air-path control system.}
    \vspace{-0.2cm}
    \label{fig:Intro-BlockHighLevel-MIMO}
    \end{figure}
   
Fig.~\ref{fig:Intro-BlockHighLevel-MIMO} shows the layout of the MIMO FCS air-path control system presented in this paper.
The signals are defined as follows: the measured outputs ($y$) are compressor mass flow $W_{cp}$, and cathode inlet/supply manifold pressure $p_{sm}$. The desired or requested stack current is $I_{st,des}$. Driver demand is mapped to this current request by a high level controller not shown in the figure. Setpoints ($r$) are the desired values of $y$, denoted by $W^*_{cp,des}$ and $p^*_{sm,des}$, and are computed based on $I_{st,des}$ using a set-point map. To maintain system constraints, an RG modifies the desired stack current to become the applied stack current $I_{st}$ and the desired setpoints $r$ to become the applied setpoints $v$. The control inputs ($u$) are the compressor motor voltage  $v_{cm}$, and the cathode outlet manifold electric throttle body (ETB) valve setting $u_{om}$. Fuel cell states are represented as $x$, which are discussed in Section \ref{section:model-O3} and can be a mix of measured and estimated values. The controller states are denoted $x_c$, which are assumed to be always known.

\subsection{Ninth-Order Plant Model}
\label{section:model-O9}

The main plant model employed in this study is a ninth-order nonlinear model of a PEM-FCS, originally developed in \cite{pukrushpan_control_2004} and agrees well with the behavior of fuel cell systems in other literature. This paper is primarily concerned with flow and pressure dynamics, so thermal and humidity subsystems need not be highly accurate, so long as the model produces valid flow and pressure predictions (e.g., mass being consumed/generated in reactions).

The original model is modified by adding the outlet manifold electronic throttle body (ETB) setting $u_{om} \in [0,1]$ as a control input. This valve is modeled as an orifice with a variable effective area, where $u_{om} = 0$ is fully closed and $u_{om} = 1$ is fully open.
{ In this simplified orifice model, we ignore the valve dynamics. This is a reasonable assumption since the closed-loop bandwidth that we seek (around 1 rad/s) is much lower than typical ETB dynamics.}

\begin{table}
    \centering
    \begin{tabular}{|c|p{0.225\linewidth}|p{0.07\linewidth}|p{0.10\linewidth}|r|r|}
        Variable & Description & Units & Baseline & Min. & Max. \\
        \hline
        $I_{st}$ & Stack Current & A & 100 & 75 & 212.5 \\
        $v_{cm}$ & Compressor Motor Voltage & V & 99 &  - & - \\
        $u_{om}$ & Outlet Manifold ETB Opening & $\frac{m^2}{m^2}$ & 0.5 & - & - \\
    \end{tabular}
    \caption{Open-loop input values for FCS model analysis.}
    \label{tab:model-op-variation}
\end{table}
To determine the degree of nonlinearity in the plant model, the ninth-order nonlinear model is linearized at multiple operating points, and the transfer functions at each operating point are compared. The baseline and variation of operating points presented in this paper are disclosed in Table~\ref{tab:model-op-variation}. All system inputs except for the one being varied were set to the baseline values.

\begin{figure}
\centering
\subfloat{\includegraphics[width=.50\linewidth] {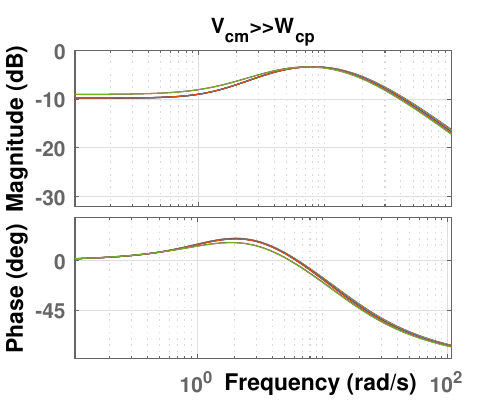}}
\subfloat{\includegraphics[width=.54\linewidth] {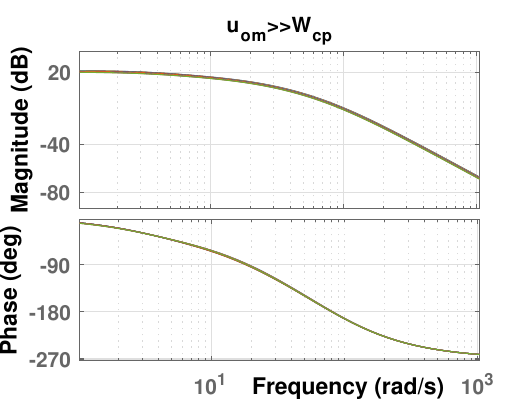}}
\caption{Superimposed bode plots of transfer functions of ninth-order model from compressor voltage { (V)} and throttle opening { (unitless)} to compressor mass airflow { (g/s)} linearized at various stack currents, evenly spaced by 12.5 A.}
\label{fig:Model-bode-u1-y4}
\end{figure}

\begin{figure}
\centering
\subfloat{\includegraphics[width=.51\linewidth] {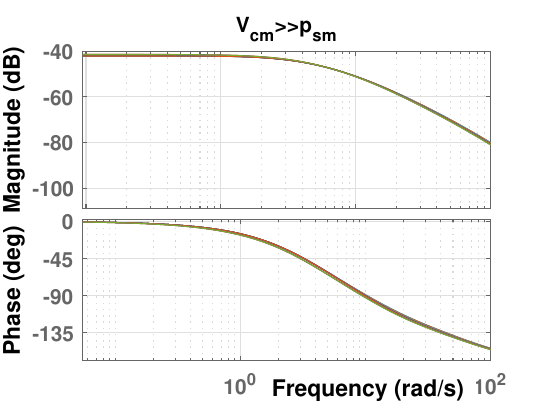}}
\subfloat{\includegraphics[width=.52\linewidth] {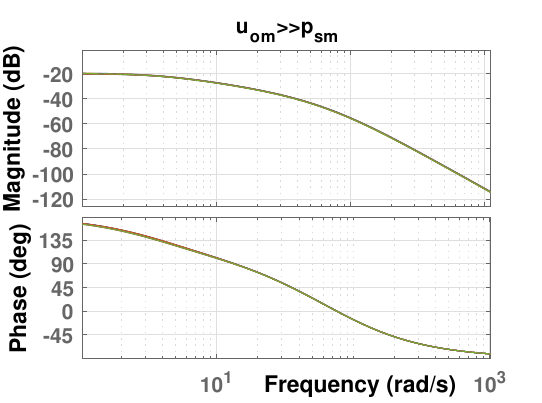}}
\caption{Superimposed bode plots of transfer functions of ninth-order model from compressor voltage {(V)} and throttle opening { (unitless)
} to cathode supply manifold pressure {(bar)} linearized at various stack currents, evenly spaced by 12.5 A.}
\label{fig:Model-bode-u1-y5}
\end{figure}

Both actuators showed almost linear input-output behavior. The nonlinearities in the fuel cell stack are not dominant, as shown in Fig.~\ref{fig:Model-bode-u1-y4} and Fig.~\ref{fig:Model-bode-u1-y5}, which show less than 10 dB magnitude variation and less than 10 degrees of phase variation. This justified the use of a single LTI feedback controller without the need for gain scheduling, at least for a system where this model's assumptions are valid (which we expect to be the case for the hardware setup that will be analyzed in our future work).

\subsection{High-Fidelity Model}
\label{section:model-hf}

In addition to the ninth-order model described in Section \ref{section:model-O9}, the control techniques developed in this project were adapted to a proprietary high-fidelity FCS model at Ford Motor Company in Section \ref{section:highfidelity}. This high-fidelity model accounted for more complex physics in the fuel cell and auxiliary components, especially for dynamics outside the air-path subsystem. This model simulated the hydration and thermal dynamics inside the fuel cell stack, and included a high-order model of the humidifier and the coolant loops. It also included closed-loop controllers for the subsystems outside the air-path, including power request tracking. This model did not include a battery or external power source, so the electric compressor was powered by the fuel cell stack.

\subsection{Third-Order Control-Oriented Model}
\label{section:model-O3}

The simplified third-order model developed and validated in \cite{Reine_J_Talj_2010} shall be used for controller design. This third-order model suitably describes the dynamic behavior of compressor mass flow and cathode supply manifold pressure. This model is augmented with an ETB at the outlet manifold, as with the ninth-order model. For the sake of completeness, we present the third-order model below (please refer to \cite{pukrushpan_control_2004} for the ninth-order model):
\begin{equation}
    \dot{p}_{ca} = \mu_2 p_{sm} - \mu_2 p_{ca} + (\mu_2 - \mu_1) p_{ca} u_{om} + \mu_3 u_{om} - \mu_4 I_{st} 
    \label{eqn:model-O3-p-ca}
\end{equation}
\begin{equation}
\begin{split}
    \dot{\omega}_{cp} = &  c_{13} v_{cm} -c_9 \omega_{cp}  \\
    & - \frac{c_{10}}{\omega_{cp}}\left[ \left( \frac{p_{sm}}{c_{11}} \right)^{c_{12}} - 1 \right] W_{cp}(\omega_{cp},p_{sm})
\end{split}
    \label{eqn:model-O3-omega-cp}
\end{equation}
\begin{equation}
\begin{split}
    \dot{p}_{sm} = & c_{14}\left[ 1 + c_{15} \left[ \left( \frac{p_{sm}}{c_{11}} \right)^{c_{12}} - 1 \right] \right] \\
    & \cdot \left[ W_{cp}(\omega_{cp},p_{sm}) - c_{16}(p_{sm} - p_{ca}) \right]
\end{split}
    \label{eqn:model-O3-p-sm}
\end{equation}

\noindent
where the state vector $x=\left[p_{ca},\omega_{cp},p_{sm}\right]^T$ consists of the air pressure in the cathode, the angular velocity of the compressor, and the pressure of air in supply manifold, respectively. $W_{cp}(\omega_{cp},p_{sm})$ is the compressor flow rate, calculated using the Jenson \& Kristensen method\cite{pukrushpan_control_2004}. The $\mu_{i}$ and $c_{i}$ constants are defined in \cite{Reine_J_Talj_2010} in terms of the nominal parameters of the FCS. Here, the parameter values were set to those in \cite{pukrushpan_control_2004} so that the parameters of the third-order model agreed with those of the ninth-order model.

Finally, the third-order nonlinear model is linearized about a single operating point. The selected point is $I^*_{st} = 190\;A$, $v^*_{cm} = 180\;V$, $u^*_{OM} = 0.45$\footnote{The maximum effective area of the valve/ETB while fully open has been defined as twice the area of the unactuated orifice at the exit of the cathode outlet manifold in \cite{pukrushpan_control_2004}. This change was made in order to track lower pressure set-points during low stack current loads as required by the new set-point maps described in Section \ref{section:set-point-maps}. For clarity, the ETB baseline value shown in Table~\ref{tab:model-op-variation} is in terms of the final enlarged orifice area.}. The ambient temperature and pressure are fixed at 298 K and 1 atm, respectively. The linearized form of the system can be described in the transfer matrix form as:
\begin{equation}
    \Delta y(s) = G(s)\Delta u(s) + G_d(s)\Delta I_{st}(s)
    \label{eqn:IMC-transfer-matrix}
\end{equation}

\noindent
where $\Delta$ stands for the deviation from the operating point (capital for s-domain), $G(s)$ is the plant, and $G_d(s)$ describes how the disturbance (i.e., stack current in this model) affects the system. For a fuel-cell system with parameters given in \cite{pukrushpan_control_2004}, $G(s)$ is obtained as follows:
\begin{equation}
\begin{aligned}
    G(s) = & \frac{1}{s^3 + 80.13s^2+1327s+2813} \\
    & \begin{bmatrix}
    20.26s^2+1171s+1061 & 23267s + 170261 \\
    1.024s + 40.38 & -174.6s - 2902
    \end{bmatrix}
\end{aligned}
    \label{eqn:IMC-plant-u-matrix}
\end{equation}

\noindent
$G(s)$ is later used in Section \ref{section:IMC} to design the air-path controller. However, $G_d(s)$ was not used in our controller design {due to the negligible effect of stack current on air supply outputs in comparison to air supply inputs as we have shown in Section \ref{section:model-O9}}, so it is not presented here.

\subsection{Relative Gain Array (RGA) Analysis}

\begin{figure}
\vspace{-0.2cm}
\centerline{\includegraphics[width=0.9\linewidth]{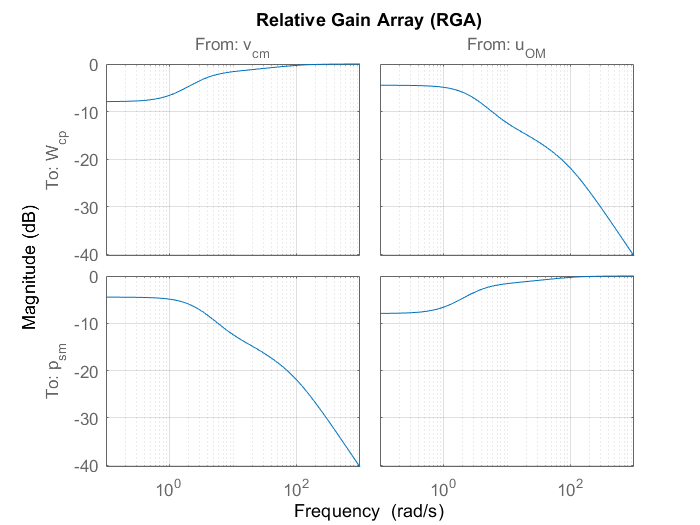}}
\vspace{-0.2cm}
\caption{RGA magnitude of $G(s)$ showing degree of input-output interaction for varying frequencies.}
\label{fig:Model-O3-RGA}
\end{figure}

The next task is to analyze the interaction between the input and output channels of the FCS air-path, before designing a MIMO controller. To determine if a diagonal controller (i.e., a separate SISO controller for each channel) can be designed for this system, a relative gain array (RGA) analysis \cite{skogestad_multivariable_1996} is conducted on the third-order transfer matrix $G(s)$ shown in (\ref{eqn:IMC-plant-u-matrix}). The RGA is obtained by:

\begin{equation}
    R(s) = G(s) \circ G(s)^{-T}
    \label{eqn:Model-RGA}
\end{equation}

\noindent
where $\circ$ is the elementwise multiplication operator. The $R(s)$ matrix describes the degree of interaction between plant channels, which informs the decision as to which outputs become feedback for each input. The magnitude of the RGA is shown in Fig.~\ref{fig:Model-O3-RGA}. According to the RGA, a diagonal controller would perform best at high frequencies, while an antidiagonal controller would perform better at low frequencies and steady-state. Therefore, to control $W_{cp}$ and $p_{sm}$ independently, two parallel SISO controllers would not suffice over the full frequency range. Motivated by this analysis, a MIMO controller is designed in Section \ref{section:IMC}.

\section{Set-Point Maps}
\label{section:set-point-maps}

This section shows how the set-point map (see Fig.~\ref{fig:Intro-BlockHighLevel-MIMO}) computes the desired compressor mass flow and desired supply manifold (or cathode inlet) pressure $r=(W^*_{cp,des},p^*_{sm,des})$.
The goal is to select set-points that maximize system net efficiency at steady-state. The compressor mass flow set-point is generated through a linear static feed-forward map (sFF) as a function of stack current. This map is analytically derived to maintain the OER at 2.0 at steady-state, similar to the method in \cite{pukrushpan_control_2004-1}. A desired OER of 2.0 was chosen based on the work of \cite{pukrushpan_control_2004}, which found this value to maximize net power output (FC stack power produced minus compressor power consumed) over a range of stack currents from 100 A to 280 A. The OER is defined as follows:
\begin{equation}
    \lambda_{O2} = \frac{W_{O2,ca,in}}{W_{O2,ca,rct}}
    \label{eqn:OER}
\end{equation}
\noindent
where
\begin{equation}
W_{O2,ca,in} = k_{a,ca,in}\frac{x_{O2,atm}}{1+\Omega_{atm}}(p_{sm} - p_{ca})
\label{eqn:W-ca-o2-in}
\end{equation}
\begin{equation}
W_{O2,ca,rct} = \frac{N_{FC}M_{O2}}{4F}I_{st}
\label{eqn:W-ca-o2-rct}
\end{equation}
\noindent
in which $W_{O2,ca,in}$ is the oxygen mass flow into the cathode, and $W_{O2,ca,rct}$ is the rate of oxygen being reacted in the cathode, $N_{FC}$ is the number of fuel cells in the stack, $M_{O2}$ is the oxygen molar mass (32 g/mol), $F$ is Faraday's constant (96485 C/mol), $k_{a,ca,in}$ is the orifice constant for the cathode inlet (defined for dry atmospheric air), $\Omega_{atm}$ is the ambient absolute humidity (1.31\%, based on relative humidity of 25\% at 298 K and 1 atm), and $x_{O2,atm}$ is the ambient/atmospheric oxygen mass fraction (0.23, based on oxygen mole fraction of 0.21). Based on (\ref{eqn:OER}), (\ref{eqn:W-ca-o2-in}) and (\ref{eqn:W-ca-o2-rct}), the compressor mass flow sFF map is:
\begin{equation}
    W^*_{cp{,des}} = \frac{1+\Omega_{atm}}{x_{O2,atm}}\frac{N_{FC}M_{O2}}{4F}\lambda^*_{O2}I_{st}
    \label{eqn:sFF-Wcp}
\end{equation}

\noindent
where $\lambda^*_{O2} = 2$.

In this study, a new sFF map is generated to map a stack current load to a supply manifold pressure set-point. The purpose of this map is to directly maximize the system net efficiency, which is defined in this study as:
\begin{equation}
    \eta_{FCS} = \frac{P_{net}}{EI_{st}} = \frac{P_{st}-P_{cp}}{EI_{st}}=\frac{v_{st}I_{st}-v_{cm}I_{cm}}{EI_{st}}
    \label{eqn:eta-fcs}
\end{equation}

\noindent
where $v_{st}$ is the PEMFC stack voltage, $I_{cm}$ is the compressor motor current, $E$ is the thermodynamic voltage of the fuel cell reaction, representing an ``ideal" reaction based on the concentrations and thermodynamic states of reactants and products, and without voltage losses (e.g., activation, ohmic, concentration, described in \cite{pukrushpan_control_2004}). This map is based on steady-state FCS efficiency, so it can be generated with any pressure controller capable of asymptotic reference tracking (i.e., through integral action).

\begin{figure}
\centerline{\includegraphics[width=0.9\linewidth]{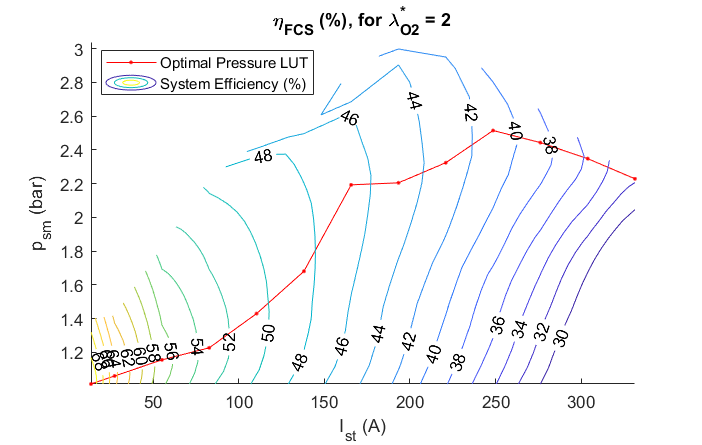}}
\vspace{-0.2cm}
\caption{Numerically generated FCS net efficiency map with optimal pressures shown.}
\label{fig:sFF-psm-LUT}
\end{figure}

To generate the sFF map, the nonlinear model with the IMC air-path controller (see {Section} \ref{section:IMC}) tracked various pressure set-points at various stack current requests. In this process, the airflow request is held constant per the discussion above to maintain $\lambda_{O2}=2$. The pressure set-point that resulted in the highest steady-state system net efficiency for each stack current was saved in a lookup-table (LUT). With these pressure values as initial guesses, gradient-based local optimization is used to refine the LUT. The final LUT is shown in Fig.~\ref{fig:sFF-psm-LUT}. This LUT is used in future simulations to generate the pressure set-point given a stack current request.

\section{Internal Model Control (IMC)}
\label{section:IMC}

To handle the challenge of controlling the PEMFC air supply system with two actuators, a multivariable internal model control (IMC) strategy is proposed to track the mass airflow and cathode inlet pressure set-points designed in Section \ref{section:set-point-maps}. IMC is straightforward to design and calibrate, even for a MIMO system with coupled dynamics. It has built-in integral action and can be made robust to plant-model mismatch by selecting sufficiently large tuning parameters to slow down the controller. These factors have made IMC an appealing control technique \cite{garcia_internal_1985,ossareh_internal_2021}.

In this section, an IMC controller is designed to compute the compressor motor input voltage $v_{cm}$ and outlet manifold ETB opening $u_{om}$ to track the desired flow and pressure set-points designed in Section \ref{section:set-point-maps}.

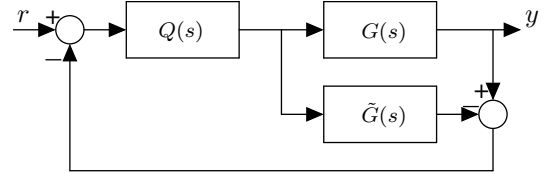
\begin{figure}
\centering
\definecolor{ffqqqq}{rgb}{1,0,0}
\definecolor{rvwvcq}{rgb}{0.0,0.0,0.00}
\begin{tikzpicture}[line width=0.5,line cap=round,line join=round,>=triangle 45, scale = 0.75]
\clip(-5,-3) rectangle (5,1);
\draw [,color=rvwvcq] (1,1)-- (3,1);
\draw [,color=rvwvcq] (3,1)-- (3,0);
\draw [,color=rvwvcq] (3,0)-- (1,0);
\draw [,color=rvwvcq] (1,0)-- (1,1);
\draw (1.5,0.8) node[anchor=north west] {{\footnotesize $G(s)$}};
\draw [,color=rvwvcq] (-2.5,1)-- (-0.5,1);
\draw [,color=rvwvcq] (-0.5,1)-- (-0.5,0);
\draw [,color=rvwvcq] (-0.5,0)-- (-2.5,0);
\draw [,color=rvwvcq] (-2.5,0)-- (-2.5,1);
\draw (-2.1,0.82) node[anchor=north west] {{\footnotesize $Q(s)$}};
\draw (1.5,-0.62) node[anchor=north west] {{\footnotesize $\Tilde{G}(s)$}};

\draw [,color=rvwvcq] (1,-0.5)-- (1,-1.5);
\draw [,color=rvwvcq] (1,-1.5)-- (3,-1.5);
\draw [,color=rvwvcq] (3,-1.5)-- (3,-0.5);
\draw [,color=rvwvcq] (1,-0.5)-- (3,-0.5);

\draw [] (-3.5,0.5) circle (0.25cm);  
\draw [->] (-3.25,0.5) -- (-2.5,0.5);    
\draw [] (0.25,0.5)-- (0.25,-1);   
\draw [->,] (0.25,-1) -- (1,-1);      
\draw [->,] (-0.5,0.5) -- (1,0.5);      

\draw [->, ] (-3.5,-2) -- (-3.5,0.25);  
\draw [->,] (-4.5,0.5) -- (-3.75,0.5);  
\draw [->,] (3,0.5) -- (4.5,0.5);  
\draw [->,] (3,-1) -- (3.75,-1);  

\draw [] (4,-1) circle (0.25cm);  

\draw (3.5,-0.3) node[anchor=north west] {{+}};  
\draw (3.3,-0.63) node[anchor=north west] {{\large\_}};  

\draw (-4.1,0.2) node[anchor=north west] {{\large\_}};  
\draw (-4.1,1.0) node[anchor=north west] {{+}};  

\draw [->, ] (4,0.5)-- (4,-0.75);         
\draw [] (4,-1.25)-- (4,-2);         
\draw [] (-3.5,-2)-- (4,-2);
\draw (-4.6,1) node[anchor=north west] {$r$};
\draw (4.4,1) node[anchor=north west] {$y$};


\end{tikzpicture}
\caption{Block diagram of the IMC architecture.}
\label{fig:IMC-block}
\end{figure}

The IMC structure is shown in Fig.~\ref{fig:IMC-block}, where  $\Tilde{G}(s)$ is the model of the plant and $Q(s)$ is the transfer matrix of the controller. Following the steps of IMC controller design from \cite{garcia_internal_1985}, the plant model can be decomposed into:

\begin{equation}
    \Tilde{G}(s) = \Tilde{G}_+(s)\Tilde{G}_-(s)
    \label{eqn:IMC-decomp}
\end{equation}

\noindent
where $\Tilde{G}_+(s)$ contains all the delays and transmission zeros of $\Tilde{G}(s)$, and $\Tilde{G}_-(s)$ contains the remaining dynamics. Since for our application $\text{det}\; \Tilde{G}(s)=0$ has no roots in the complex right half-plane, the inverse plant model has no nonminimum phase zeros and the factorization matrix can be selected as $\Tilde{G}_+(s) = I$ which yields $Q(s) = \Tilde{G}^{-1}(s)$. But the inversion of the $\Tilde{G}(s)$ transfer matrix in (\ref{eqn:IMC-plant-u-matrix}) is still an improper transfer matrix. To obtain a realizable controller, it is augmented with a low pass filter (as it is done in the standard IMC design):

\begin{equation}
    Q(s) = \Tilde{G}^{-1}(s)F(s)
    \label{eqn:IMC-controller}
\end{equation}

\noindent
where we select the following structure for the low-pass filter transfer matrix:
\begin{equation}
    F(s) = \begin{bmatrix}
    \frac{1}{(\tau_1 s + 1)^{n_1}} & 0 \\
    0 & \frac{1}{(\tau_2 s + 1)^{n_2}}
    \end{bmatrix}
    \label{eqn:IMC-filter}
\end{equation}
In an effort towards decoupling the two reference signals, a diagonal structure was chosen for the filter. The values of $n_1=1$ and $n_2=2$ are used, as these are the minimal values that can make the controller transfer function proper. The time constants of the low pass filters ($\tau_1$ and $\tau_2$) are used as tuning parameters to trade-off between performance and robustness. Here, the time constants are selected as $\tau_1=0.20$ and $\tau_2=0.20$ seconds.

Table~\ref{tab:IMC-margins} shows the loop-at-a-time and multiloop disk margins for variations at the plant input. A review on disk margins can be found in \cite{SeilerPeter2020AItD}. When compared to the plant's degree of nonlinearity characterized in Section \ref{section:model}, the maximum gain and phase variations fit within the disk-based margins, guaranteeing closed-loop stability over a wide range of stack currents. It can be concluded that the MIMO air-path IMC controller guarantees robustness when there is uncertainty in one of the actuators or both at the same time.

\begin{table}
\caption{MIMO Air-Path IMC Disk-Based Stability Margins}
\begin{center}
\begin{tabular}{|c|c|c|c|}
\hline
\cline{2-4} 
\textbf{Channel: } & \textbf{$\mathbf{v_{cm}}$}& \textbf{$u_{om}$}& {Both} \\
\hline
Gain Margin [dB] & $\pm 20.38$ & $\pm 17.14 $ & $\pm 12.11$  \\
\hline
Phase Margin & $\pm 79.07^{\circ}$ & $\pm 74.17^{\circ}$ & $\pm 62.15^{\circ} $  \\
\hline
\end{tabular}
\label{tab:IMC-margins}
\end{center}
\end{table}

\begin{figure}
\centerline{\includegraphics[width=1.0\linewidth]{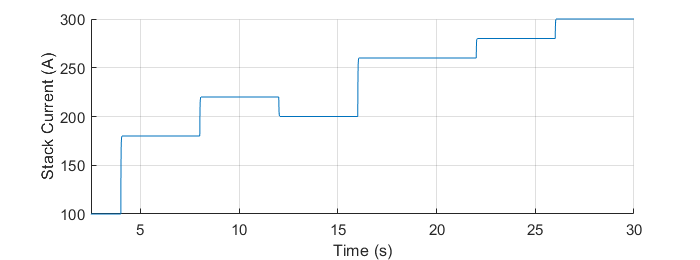}}
\caption{Stack current steps.}
\label{fig:IMC-Ist}
\end{figure}

The MIMO air-path IMC is tested on the nonlinear full-state (ninth-order) FCS model, using a series of stack current steps shown in Fig.~\ref{fig:IMC-Ist}. The stack current sequence was based on \cite{pukrushpan_control_2004}. A 10 ms actuator delay was added to the simulation to further test the controller's robustness.
The flow and pressure set-points are computed with sFF maps based on stack current as described in Section \ref{section:set-point-maps}.

\begin{figure}
\centerline{\includegraphics[width=0.9\linewidth]{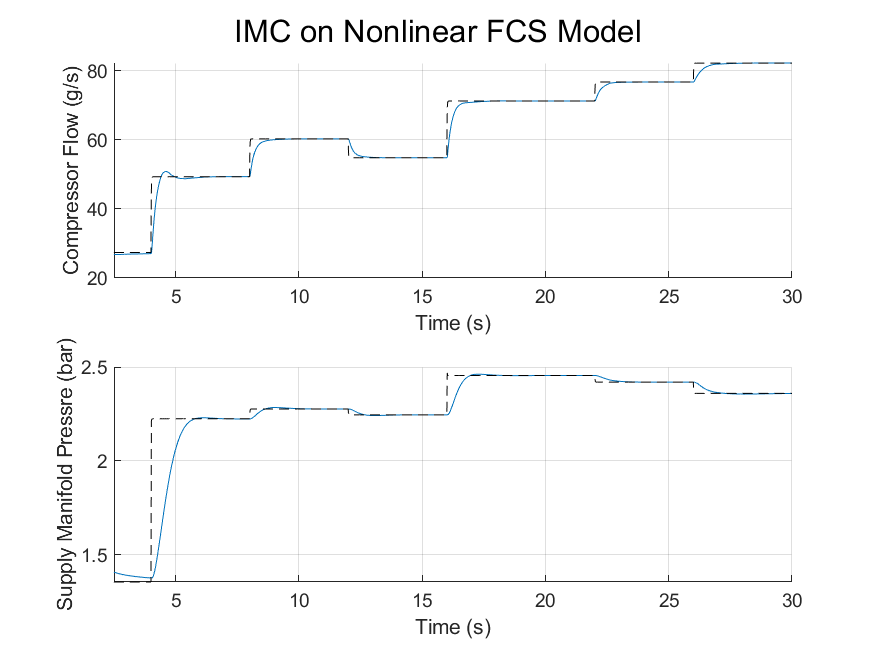}}
\caption{Air-path output response with MIMO IMC on the ninth-order nonlinear model.}
\label{fig:IMC-y}
\end{figure}

The compressor mass flow and cathode inlet pressure responses of the MIMO air-path IMC are shown in Fig.~\ref{fig:IMC-y}. Both responses show little to no overshoot and rise-time in the one-second range.

\begin{figure}
\centerline{\includegraphics[width=0.9\linewidth]{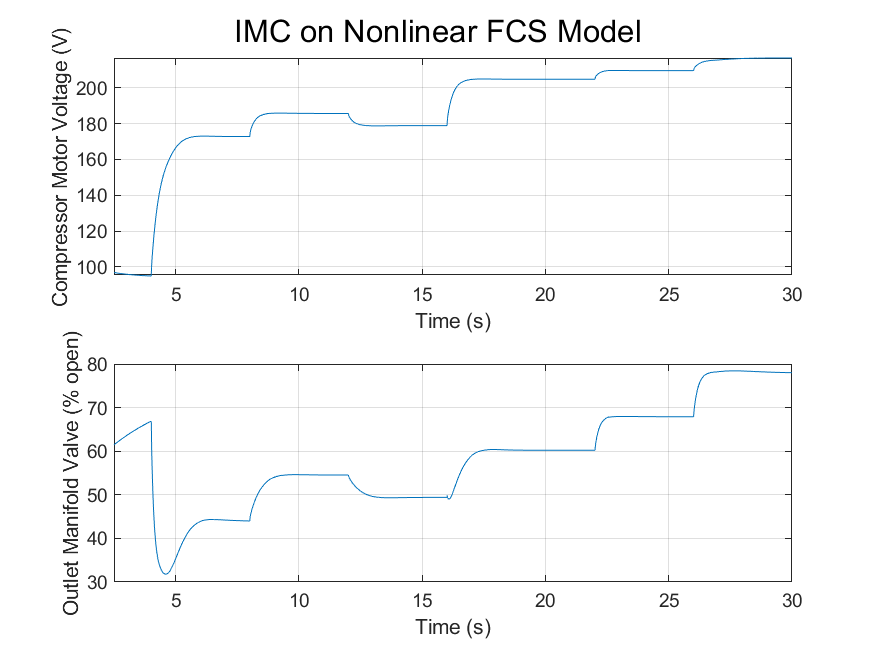}}
\caption{Air-path actuator commands from MIMO IMC on the ninth-order nonlinear model.}
\label{fig:IMC-u}
\end{figure}

The actuator plots for the MIMO air-path IMC are shown in Fig.~\ref{fig:IMC-u} for the compressor motor voltage and cathode outlet manifold ETB. These plots show stable and non-oscillatory actuator behavior. The non-minimum phase ETB response indicates that the ETB is first closed to increase the pressure and then opened to increase the flow. This demonstrates how the MIMO air-path IMC handles the  coupling between the two actuators and the two outputs shown in Fig.~\ref{fig:Model-O3-RGA}.

\begin{figure}
\centerline{\includegraphics[width=0.9\linewidth]{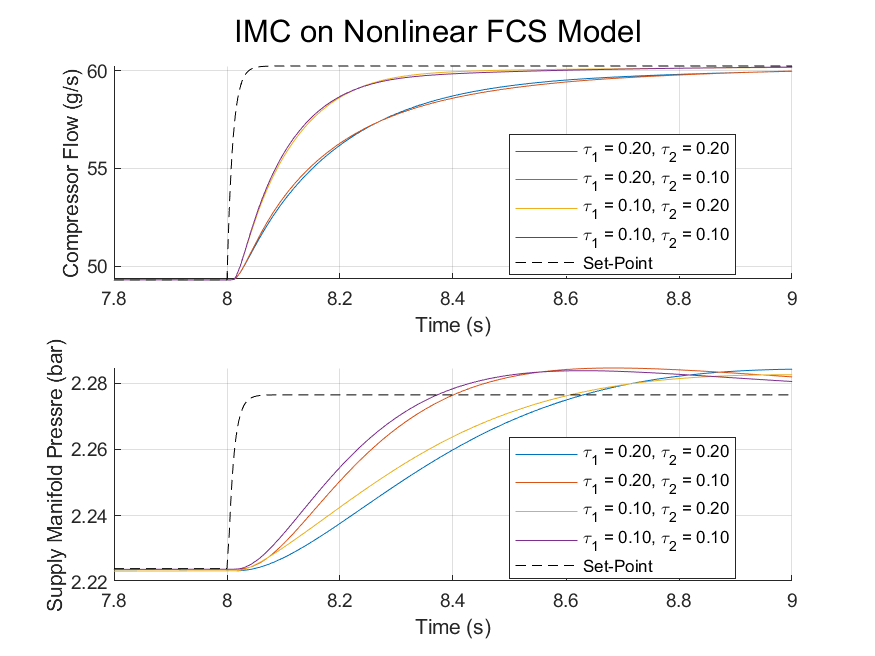}}
\caption{Comparisons of tuning $\tau$ on MIMO IMC on the ninth-order nonlinear model.}
\label{fig:IMC-tune}
\end{figure}

Calibrating the IMC for a desired trade-off between speed and robustness is a straightforward process. Each time constant corresponds to one of the two references: $\tau_1$ corresponds to compressor flow, while $\tau_2$ corresponds to supply manifold pressure. The calibration process of $\tau_1$ and $\tau_2$ is shown in Fig.~\ref{fig:IMC-tune}. Decreasing either $\tau$ increases tracking speed for the corresponding reference signal, while increasing either $\tau$ slows down the tracking to increase robustness. This intuitive calibration makes the IMC easy to modify to suit specific system parameters or different control objectives.

\section{Background and Baseline FCS Application of Reference Governors (RG)}
\label{section:RG}

\begin{figure}
    \centering
    \begin{tikzpicture}[auto, node distance=1.5cm,>=latex']
    \node [input, name=rinput] (rinput) {};
    \node [block, right of=rinput,text width=1.5cm,align=center] (controller) {{\footnotesize Reference Governor}};
    \node [block, right of=controller,node distance=3cm,text width=1.6cm,align=center] (system)
    {{\footnotesize Closed-Loop System}};
    \node [output, right of=system, node distance=2cm] (output) {};
    \node [tmp, below of=controller,node distance=0.9cm] (tmp1){$s$};
    \draw [->] (rinput) -- node{\hspace{-0.4cm}$r(t)$} (controller);
    \draw [->] (controller) -- node [name=v]{$v(t)$}(system);
    \draw [->] (system) -- node [name=y] {$y(t)$}(output);
    \draw [->] (system) |- (tmp1)-| node[pos=0.75] {$x(t)$} (controller);
    \end{tikzpicture}
    \caption{Reference governor block diagram.}
    \label{fig:RGblock}
\end{figure}
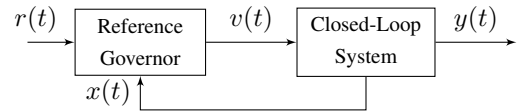

Next, a reference governor (RG) is implemented on the FCS to maintain a minimum OER constraint at all times.
Compressor surge constraints can also be enforced using this same methodology, but surge constraint management is not presented here because surge was not an issue with our compressor.

An RG is a predictive constraint management technique that modifies the input (e.g., reference) signal to a closed-loop control system to maintain output constraints in both transient and steady-state. Using measurements or estimates of all states describing the closed-loop system at each timestep, the RG modifies the reference as little as possible in order to maintain constraints. This setup is shown in Fig.~\ref{fig:RGblock}. {If the RG in this setup employs a linear prediction model, it would have a closed-form solution and thus be computationally efficient}.
A brief overview of RG theory is presented in Section \ref{section:RG-theory}.

The RG has been considered for FCS applications in prior literature. An RG to prevent oxygen starvation using a nonlinear prediction model is studied in \cite{jing_sun_load_2005}. This implementation involved running a nonlinear FCS model and bisectional search algorithm on-line. In \cite{vahidi_constraint_2007}, an RG using a linear prediction model, as well as a disturbance term to handle plant-model mismatch, was applied to enforce OER constraints and compressor surge constraints. In \cite{vahidi_constraint_2005}, a linear prediction model RG is compared against an MPC to enforce OER and compressor surge constraints, showing comparable performance between the two methods, but a smaller average number of floating point operators for the RG.

All three of the above sources were ultimately successful in maintaining OER constraints by governing the stack current load. However, these prior cases considered a SISO air-path control system, which controlled mass airflow with compressor motor voltage. Our paper extends their results to function on our MIMO air-path control system (i.e., including the ETB actuator and pressure control) in Section \ref{section:RGi}. This RG is similar to \cite{vahidi_constraint_2007}, but applied to the MIMO closed-loop air-path. There are two more key differences: first, compressor surge constraints are not included in this work, as surge was not an issue with our compressor model. Second, we did not adopt the disturbance term to handle modeling error, which is instead mitigated with a newly reformulated OER constraint, which will be described in \ref{section:RG-OER}.

While the {load governor} presented in Section \ref{section:RGi} successfully maintains the OER constraint, it can still significantly slow down the load request, slowing down the power dynamics, a behavior shared by other FCS load governors such as those in \cite{jing_sun_load_2005,vahidi_constraint_2007}. This shortcoming motivates the second, novel RG formulation presented in Section \ref{section:RGap}, to maintain safe minimum OER while also more closely following the ungoverned net power delivery. The novel RG in Section \ref{section:RGap} is based on two new theoretical frameworks, which will be presented in this paper and may have applications beyond FCS. The performance of the second RG is then compared to the first more traditional { single-reference, variable time-constant lowpass filter, linear prediction model} RG, which serves as a baseline.

\subsection{General RG Theory}
\label{section:RG-theory}

Again consider Fig.~\ref{fig:RGblock}, in which the ``closed-loop system'' is described by the single-input multi-output discrete-time, stable linear system: 
\begin{equation}
\label{eqn:system_cont}
\begin{aligned}
    x[k+1] = Ax[k]+Bv[k]\\
    y[k] = Cx[k]+Dv[k]
\end{aligned}
\end{equation}
with the output $y$ subject to the following polyhedral constraints:
\begin{equation}\label{eqn:Sys}
    y[k] \in \mathbb{Y} \triangleq \{y[k]: S y[k+j]\leq s\}, \forall j \in \mathbb{Z}_{+}
\end{equation}

To clarify, square brackets $[\cdot]$ are used here to denote discrete-time signals (i.e., sampled from { the} continuous-time system), such that $x[k]=x(kT_s)$ where $T_s$ is the sampling period and $k$ is the timestep. The state $x\in\mathbb{R}^n$ and output $y\in\mathbb{R}^p$ are vectors, and the input to the closed-loop system $v$ is a scalar (this assumption will later be relaxed in Section \ref{section:RG-cascade} when governing multiple inputs). The constraint set is defined by $s\in\mathbb{R}^{q}$ and $S\in\mathbb{R}^{q\times p}$, where $q$ is the number of constraints.

To predict and prevent constraint violation, the RG employs the so-called maximal admissible set (MAS), denoted by $O_\infty$ \cite{Gilbert_1991}. The MAS is the set of all states $x$ and constant inputs $v$ that satisfy \eqref{eqn:Sys} for all time:
\begin{equation}\label{eqn:Oinfdefintiion}
O_\infty = \big\{(x,v): x[0]=x,\ v[k]=v,\ y[k] \in \mathbb{Y},\ \forall k\in \mathbb{Z}^+\big\}
\end{equation}
\noindent
where $y[k]$ is the predicted output given by:
\begin{equation}
    y[k] = CA^k x[0] + \left(C(I-A^k)(I-A)^{-1}B+D\right) v
    \label{eqn:y[k]}
\end{equation}

It is shown in \cite{Gilbert_1991} that, under mild assumptions on $C$ and $A$, it is possible to make this set finitely determined by constraining the steady-state value of $y$. This leads to an inner approximation of MAS that can be computed offline: 
\begin{equation}
   \widetilde{O}_{\infty} = \Big\{(x,v):H_x x + H_v v \leq h\Big\}
   \label{eqn:O_infty}
\end{equation}
where the matrices $H_x$, $H_v$, and $h$ are finite-dimensional { with number of rows $k^*$}, and are defined { by predicting the system's evolution for each future time-step within a finite horizon with} (\ref{eqn:Sys}) and (\ref{eqn:y[k]}).
The RG leverages $\widetilde{O}_\infty$ to select a $v$ at every timestep to enforce the constraints. The RG update law that achieves this is:
\begin{equation}
   v[k] = v[k-1]+\kappa [k] \left(r[k]-v[k-1]\right)
   \label{eqn:vkappa}
\end{equation}
where $\kappa \in [0,1]$, calculated via an online linear program (LP):
\begin{equation}
    \begin{aligned}
    &\underset{\kappa\in [0,1]}{\text{maximize}}
    & & \mathrm{\kappa} \\
    & \hspace{10pt} \text{s.t.}
    & & v[k]=v[k-1]+\kappa\left(r[k]-v[k-1]\right)\\
    &&&\left(x[k],\ v[k]\right) \in \widetilde{O}_{\infty}
    \end{aligned}
    \label{eqn:kappa_for_RG}
\end{equation}

If $\kappa=0$, the control command from the previous timestep is maintained to avoid constraint violation. If $\kappa = 1$, $v[k] = r[k]$. The value for $\kappa$ can be solved explicitly, without the need for a LP solver. The update law shown in (\ref{eqn:vkappa}) causes the RG to behave like a low-pass filter with a variable time constant $\kappa$. For more information, please refer to \cite{Gilbert_1991, Gilbert_Kolmanovsky_1995, Ossareh_2019, liu2018decoupled}.

\subsection{RG Applied to OER}
\label{section:RG-OER}

The OER itself is a nonlinear combination of system states and inputs, as shown in (\ref{eqn:OER}). This creates an inevitable limit to the accuracy of predicting OER dynamics with a linear system model. To address this, the OER constraint is reformulated through a selection of $S$ and $s$ in (\ref{eqn:Sys}), which differs from the RG's previously applied to FCS in \cite{vahidi_constraint_2005,vahidi_current_2006,vahidi_constraint_2007}. By multiplying both sides of (\ref{eqn:OER}) by $W_{O2,ca,rct}$, the minimum OER constraint can now be formulated as:
\begin{equation}
    \lambda_{O2,min} W_{O2,ca,rct} - W_{O2,ca,in} \leq 0 
    \label{eqn:OER-lin-out}
\end{equation}
By substituting the expressions for $W_{O2,ca,in}$ and $W_{O2,ca,rct}$ in terms of system states and inputs, the constraint in (\ref{eqn:OER-lin-out}) becomes:
\begin{equation}
    \lambda_{O2,min}(\frac{N_{FC}M_{O2}}{4F}I_{st}) - (k_{a,ca,in}\frac{x_{O2,atm}}{1+\Omega_{atm}})(p_{sm} - p_{ca})  \leq 0 
    \label{eqn:OER-lin-state}
\end{equation}
Equation (\ref{eqn:OER-lin-state}) represents a linear combination of system states and inputs, although the states themselves may evolve according to nonlinear dynamics. From here, the RG's MAS can be computed offline using the following:

\begin{equation}
\begin{matrix}
y = \left( \delta W_{O2,ca,in}, \delta W_{O2,ca,rct} \right) \\
\\
S = \begin{bmatrix}
-1 & \lambda_{O2,min}
\end{bmatrix} \\
s = y^*_{1} - \lambda_{O2,min}y^*_{2}\\
\end{matrix}
\label{eqn:RG-form}
\end{equation}

\noindent
where $y_1^*$ and $y_2^*$ are the output values at the model linearization point, and $\delta(\cdot)$ representing deviation from the linearization point (lowercase for time-domain). The RG is fed the augmented state $z = \begin{bmatrix} \delta x, \delta x_{c} \end{bmatrix}^T$. The air-path controller state $x_{c}$ is directly measurable, and the plant state $x$ can be a mix of measurements and estimations based on sensor availability. RG implementations in this project use the third-order FCS model. The compressor shaft angular velocity and the cathode supply manifold pressure states are directly measured, while the cathode pressure state must be estimated. The importance of this new formulation is that it reduces a known source of plant-model mismatch for an RG (namely, the nonlinear constraint) using a computationally efficient linear prediction model.

\subsection{Load governor}
\label{section:RGi}

\begin{figure}
\centering
\includegraphics[page=7,width=0.9\linewidth]{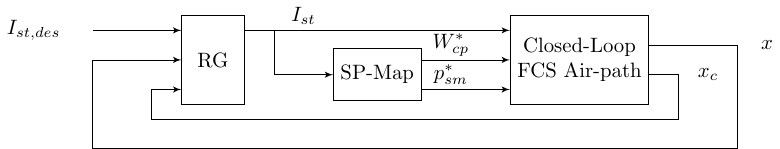}
\caption{{Load governor}. The ``SP-Map'' block is the Set-Point Map.}
\label{fig:block-RGi}
\vspace{-0.4cm}
\end{figure}

The {load governor} implemented for OER is shown in Fig.~\ref{fig:block-RGi}. Note that Fig.~\ref{fig:block-RGi} is not the same RG shown in Fig.~\ref{fig:Intro-BlockHighLevel-MIMO}, rather it serves as a baseline to compare against the novel RG presented in Section \ref{section:RGap}. The {load governor} is a straightforward application of the methods described in the literature aforementioned in Section \ref{section:RG-theory}.
While the online and offline RG algorithms are not novel, the application of a standard RG to the MIMO airpath subsystem is new, as well as the linear OER constraint formulation presented in \eqref{eqn:OER-lin-out}.

\begin{figure}
\centerline{\includegraphics[width=0.9\linewidth]{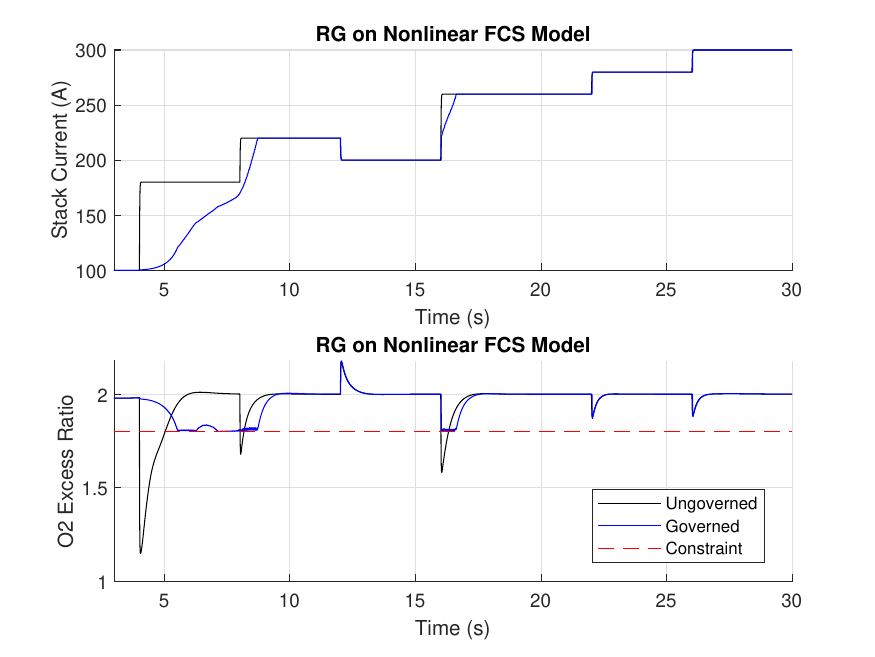}}
\caption{Stack current and OER response to current steps, without and with the load governor on the ninth-order nonlinear model.}
\label{fig:RG-OER}
\end{figure}

The system with the {load governor} and the MIMO air-path IMC was tested on the same series of stack current steps used in Section \ref{section:IMC} to maintain OER above 1.8. The OER response without and with the RG, as well as the corresponding stack current drawn from the FC in each case, are shown in Fig.~\ref{fig:RG-OER}. Unlike the ungoverned case, the governed case maintains the OER constraint at all times. When the RG is active ($\kappa<1$), the OER response reaches the constraint without violating it. This behavior occurs at both low (100 A) and medium (200 A) stack currents, indicating that the RG remains effective for a wide range of operating points. This can be attributed to the new OER constraint formulation (\ref{eqn:OER-lin-state}) being linear in terms of stack current. If the RG were instead implemented with the following more intuitive constraint,

\begin{equation}
    \lambda_{O2} = \frac{W_{O2,ca,in}}{W_{O2,ca,rct}} \leq \lambda_{O2,min}
    \label{eqn:OER-nonlin}
\end{equation}

\noindent
then $I_{st}$ would be in the denominator of (\ref{eqn:OER-nonlin}). This nonlinearity may lead to constraint violation at lower currents and overly conservative RG action at higher currents. The {load governor} does not modify the system response when there is no danger of constraint violation. Thus, the governed current converges to ungoverned current as the system approaches steady-state.

\section{Performant RG for MIMO Systems with Application to FCS}
\label{section:RGap}

The RG presented in Section \ref{section:RG} and the RGs in \cite{vahidi_constraint_2007} and \cite{jing_sun_load_2005} share the limitation of manipulating only the stack current load. This can considerably slow down fuel cell power production. This limitation led to the development of a novel mutlivariable RG that effectively enforces constraints while {retaining} the use of a computationally efficient linear prediction model. { To distinguish this from the {load governor} presented in Section \ref{section:RGi}, the novel RG will be referred to as the cascaded cross-section RG (CC-RG). Note that the proposed CC-RG architecture is mainly intended for the FCS OER application, but may be suitable for other applications as well.}

{The CC-RG contains two novel and independent RG  frameworks.} {First, the Cascade RG presented in Section \ref{section:RG-cascade} modifies multiple set-points by arranging several RG in series. Second, the Cross-Section RG (CS-RG) presented in Section \ref{section:RG-CS}, is capable of speeding up as well as slowing down a reference signal to maintain constraints. This enables} the applied reference to temporarily (and intentionally) overshoot the desired reference when needed. This generalizes the RG update law (\ref{eqn:vkappa}), which for FCS applications will prove useful for increasing oxygen supply to reduce OER drops.

\begin{figure}
    \centering
    \includegraphics[page=8,width=0.9\linewidth]{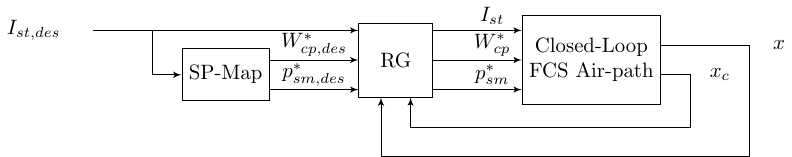}
    \caption{CC-RG.}
    \label{fig:block-RGap}
    \vspace{-0.4cm}
\end{figure}

The CC-RG {for an FCS application} is depicted at a high level in Fig.~\ref{fig:block-RGap}, which represents a ``zoomed-in'' view of the ``Reference Governor'' block in Fig.~\ref{fig:Intro-BlockHighLevel-MIMO}. This RG modifies the stack current, compressor flow set-point, and the cathode inlet pressure set-point using a Cascade RG. In the case of compressor flow, this RG has the ability to ``speed-up'' the requested flow, rather than strictly low-pass filtering the signal, using a CS-RG. The CC-RG will be tested on stack current steps in Section \ref{section:RGap-sim}, and its net power response will be compared against the ungoverned case and that of the {load governor}.

\subsection{Cascade RG}
\label{section:RG-cascade}

The Cascade RG { employs multiple RGs in series. Each RG can only govern one reference signal but is still aware of the other signals it cannot manipulate by accepting the applied/governed output of the RG preceding it.
To describe each  sub-RG within the cascade RG, we group the setpoints into two groups, $v$ and $w$, where $v$ is the reference to be governed by the sub-RG and $w$ represents either a desired set-point to be modified by a sub-RG later in the cascade or an applied set-point modified by a sub-RG earlier in the cascade. Accordingly, the MAS inner-approximation (\ref{eqn:O_infty}) can be expressed as shown below}:
\begin{equation}
    H_x x + H_v v + H_w w \leq h
    \label{eqn:RGap-MAS-cascade}
\end{equation}
where the scalar $v\in\mathbb{R}$ is the manipulated reference, the vector $w$ are the feedthrough references, $[ v \; w ]^T\in\mathbb{R}^m$ for a closed-loop system with $m$ inputs, $H_w\in\mathbb{R}^{k^* \times (m-1)}$, and $H_v\in\mathbb{R}^{k^*}$ just as in a scalar RG. The new matrix
$H_w$ for each feedthrough reference $w$ is computed by the same method as $H_v$ for manipulated reference $v$.
In order to design a cascade-RG, one can arrange each RG in series in ascending order of priority. A reference that is deemed more free to vary is governed before a reference that should be kept as close to its desired value {$r[k]$} as possible, as we will show { for the FCS application} in Section \ref{section:RGap-sim}. { Note that changing the order in which references are governed can lead to significant differences in system behavior and performance, making it an important design decision.}

{Under the assumption that each sub-RG in the cascade has a bounded output $v[k]$, which is the case when the maximal admissible set is compact, and that the system in \eqref{eqn:system_cont} is bounded-input bounded-output (BIBO) stable, a system controlled with a Cascade RG will be BIBO stable. If the sub-RGs within the cascade are scalar RGs with update law \eqref{eqn:vkappa}, then $v[k]$ also forms a monotonic sequence, which guarantees convergence to a constraint-admissible setpoint and thus asymptotic stability for a constant $r[k]$.
}

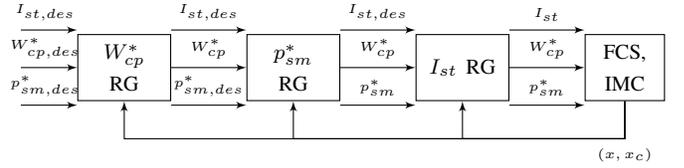
\begin{figure}
    \centering
    \begin{tikzpicture}[auto, node distance=1.5cm,>=latex']
        \node [input, name=rinput] (rinput) {};
        \node [tmp, above=0.5cm of rinput] (Istdes) {};
        \node [tmp, above=0.0cm of rinput] (Wcpdes) {};
        \node [tmp, below=0.5cm of rinput] (psmdes) {};
        
        \node [tmp, right=0.5 of Istdes] (tmpIstdes1) {};
        \node [tmp, right=0.5 of psmdes] (tmppsmdes1) {};
        
        \node [block, right=0.75cm of rinput,text width=1.0cm,align=center] (RGw) {{\footnotesize $W_{cp}^*$ RG}};
        \node [tmp, left=0cm of RGw, yshift=0.50cm] (RGw_Ii) {};
        \node [tmp, left=0cm of RGw, yshift=0.00cm] (RGw_wi) {};
        \node [tmp, left=0cm of RGw, yshift=-0.50cm] (RGw_pi) {};
        \node [tmp, right=0cm of RGw, yshift=0.50cm] (RGw_Io) {};
        \node [tmp, right=0cm of RGw, yshift=0.00cm] (RGw_wo) {};
        \node [tmp, right=0cm of RGw, yshift=-0.50cm] (RGw_po) {};

        \draw [->] (Istdes) -- node{\tiny \hspace{-0.10cm}$I_{st,des}$} (RGw_Ii);
        \draw [->] (Wcpdes) -- node{\tiny \hspace{-0.10cm}$W^*_{cp,des}$} (RGw_wi);
        \draw [->] (psmdes) -- node{\tiny \hspace{-0.10cm}$p^*_{sm,des}$} (RGw_pi);

        \node [tmp, right=0.5 of RGw_Io] (tmpIstdes2) {};
        \node [tmp, above=0.5 of tmpIstdes2] (tmpIstdes2up) {};
        \node [tmp, right=0.5 of RGw_po] (tmppsmdes2) {};
        \node [tmp, below=0.5 of tmppsmdes2] (tmppsmdes2dn) {};

        \node [block, right=1.0cm of RGw,text width=1.0cm,align=center] (RGp) {{\footnotesize $p_{sm}^*$ RG}};
        \node [tmp, left=0cm of RGp, yshift=0.50cm] (RGp_Ii) {};
        \node [tmp, left=0cm of RGp, yshift=0.00cm] (RGp_wi) {};
        \node [tmp, left=0cm of RGp, yshift=-0.50cm] (RGp_pi) {};
        \node [tmp, right=0cm of RGp, yshift=0.50cm] (RGp_Io) {};
        \node [tmp, right=0cm of RGp, yshift=0.00cm] (RGp_wo) {};
        \node [tmp, right=0cm of RGp, yshift=-0.50cm] (RGp_po) {};

        \draw [->] (RGw_Io) -- node{\tiny \hspace{-0.00cm}$I_{st,des}$} (RGp_Ii);
        \draw [->] (RGw_wo) -- node{\tiny \hspace{-0.00cm}$W^*_{cp}$} (RGp_wi);
        \draw [->] (RGw_po) -- node{\tiny \hspace{-0.00cm}$p^*_{sm,des}$} (RGp_pi);

        \node [block, right=1.0cm of RGp,text width=1.0cm,align=center] (RGi) {{\footnotesize $I_{st}$ RG}};
        \node [tmp, left=0cm of RGi, yshift=0.50cm] (RGi_Ii) {};
        \node [tmp, left=0cm of RGi, yshift=0.00cm] (RGi_wi) {};
        \node [tmp, left=0cm of RGi, yshift=-0.50cm] (RGi_pi) {};
        \node [tmp, right=0cm of RGi, yshift=0.50cm] (RGi_Io) {};
        \node [tmp, right=0cm of RGi, yshift=0.00cm] (RGi_wo) {};
        \node [tmp, right=0cm of RGi, yshift=-0.50cm] (RGi_po) {};

        \draw [->] (RGp_Io) -- node{\tiny \hspace{-0.00cm}$I_{st,des}$} (RGi_Ii);
        \draw [->] (RGp_wo) -- node{\tiny \hspace{-0.00cm}$W^*_{cp}$} (RGi_wi);
        \draw [->] (RGp_po) -- node{\tiny \hspace{-0.00cm}$p^*_{sm}$} (RGi_pi);

        \node [block, right=1.0cm of RGi,text width=0.75cm,align=center] (sys) {{\footnotesize FCS, IMC}};
        \node [tmp, left=0cm of sys, yshift=0.50cm] (sys_Ii) {};
        \node [tmp, left=0cm of sys, yshift=0.00cm] (sys_wi) {};
        \node [tmp, left=0cm of sys, yshift=-0.50cm] (sys_pi) {};

        \draw [->] (RGi_Io) -- node{\tiny \hspace{-0.00cm}$I_{st}$} (sys_Ii);
        \draw [->] (RGi_wo) -- node{\tiny \hspace{-0.00cm}$W^*_{cp}$} (sys_wi);
        \draw [->] (RGi_po) -- node{\tiny \hspace{-0.00cm}$p^*_{sm}$} (sys_pi);

        \node [tmp, below=0.5cm of RGw] (tmpRGwx) {};
        \node [tmp, below=0.5cm of RGp] (tmpRGpx) {};
        \node [tmp, below=0.5cm of RGi] (tmpRGix) {};

        \draw[->] (sys) |- node{\tiny $(x,x_c)$} (tmpRGix) -- (RGi);
        \draw[->] (sys) |- (tmpRGpx) -- (RGp);
        \draw[->] (sys) |- (tmpRGwx) -- (RGw);
    \end{tikzpicture}
    \caption{Block diagram of the cascade RG for FCS OER constraint management. Each sub-RG block has two feedthrough references and one manipulated reference.}
    \label{fig:block-RG-cascade}
    \vspace{-0.4cm}
\end{figure}

{The Cascade RG can be applied to the OER problem in an FCS.} Stack current still needs to be governed to maintain the minimum OER constraint. Upon inspection of (\ref{eqn:W-ca-o2-rct}), stack current appears as a direct-feed-through term on oxygen consumption, giving it a rapid effect on OER. But by adding degrees of freedom through the governing of desired flow and pressure, the stack current can play a smaller role in constraint management and thus still be drawn from the fuel cell at a faster rate than by the {load governor}, leading to faster power delivery. In the case of the FCS air-path, it is three RG in series: one for stack current, one for the flow set-point, and one for the pressure set-point. This is depicted in Fig.~\ref{fig:block-RG-cascade}, which shows the contents of the ``RG'' block in Fig.~\ref{fig:block-RGap}.

A potential alternative to governing multiple references in series with a cascade RG is governing multiple references in parallel, grouping by closed-loop inputs that move the system in the same direction with respect to the constraint. For example, slowing down stack current and the pressure setpoint both move the FCS away from the minimum OER constraint, so current and pressure could be coupled to one another in the RG prediction model and governed as one rather than two reference signals. This may be an interesting avenue for future research.

\subsection{Cross-Section RG (CS-RG)}
\label{section:RG-CS}

In order for the compressor mass flow reference to be useful in maintaining safe OER, the applied airflow reference must be able to temporarily overshoot the desired flow reference. This is because in order to use airflow to prevent oxygen starvation, the oxygen supply to the cathode would have to be sped-up, while a traditional RG would only be able to slow it down. This led to the development of the Cross-Section RG (CS-RG), which generalizes the online reference governor update law represented in (\ref{eqn:vkappa}). { The CS-RG was developed for the specific application of OER constraint management of PEM fuel cells, but there may be other applications for such a strategy as well.}

{ The CS-RG differs from the RG in \cite{Gilbert_1991,Gilbert_Kolmanovsky_1995} in that the applied scalar reference $v[k]$ is no longer limited to a convex combination of the $v[k-1]$ and $r[k]$. In the CS-RG, a reference signal can be increased or decreased, { and it can even overshoot or undershoot the desired $r[k]$}, such that the constraint is satisfied.
In addition, in the CS-RG $v[k]$ does not depend explicitly on $v[k-1]$.
} The offline calculations to construct the MAS remain the same as those in {Section} \ref{section:RG-theory}.

Recall that {for the LTI system in \eqref{eqn:system_cont}} a standard RG calculates the applied reference signal by bounding it within the inner approximation of MAS, $\widetilde{O}_{\infty}$. Recall also that this set is polyhedral and so it consists of the intersection of half-spaces of the form:
\begin{equation}
    H_{x,j}x + H_{v,j}v \leq h_{j}
    \label{eqn:RGap-MAS-j}
\end{equation}

\noindent
where $(\cdot)_j$ represents the $j$-th row of the matrix $(\cdot)$, corresponding to a single row of the inequalities defining $\widetilde{O}_{\infty}$ (\ref{eqn:O_infty}). There is only one unknown in this inequality, $v$, while all other values are known. Rearranging terms, we obtain:

\begin{equation}
    H_{v,j}v \leq h_{j} - H_{x,j}x
    \label{eqn:RGap-MAS-j-Hv-v}
\end{equation}

{This allows us to find lower and upper bounds on $v[k]$ at each timestep $k$ by dividing both sides of \eqref{eqn:RGap-MAS-j-Hv-v} by $H_{v,j}$. However, the direction of the inequality ($\leq$ vs. $\geq$) depends on the sign of the denominator $H_{v,j}$. We thus have:}

{
\begin{equation}
\begin{array}{ll}
    v_{max}[k] \;=\; \min\limits_{j:\;H_{v,j} > 0} \{ \frac{h_{j} - H_{x,j}x[k] }{H_{v,j}} \} \\
    v_{min}[k]  \;=\; \max\limits_{j:\;H_{v,j} < 0}\{ \frac{h_{j} - H_{x,j}x[k] }{H_{v,j}} \}
\end{array}
\label{eqn:RGap-CS-v}
\end{equation}
}
{

\begin{figure}
\vspace{-0.4cm}
\centerline{\includegraphics[width=0.6\linewidth]{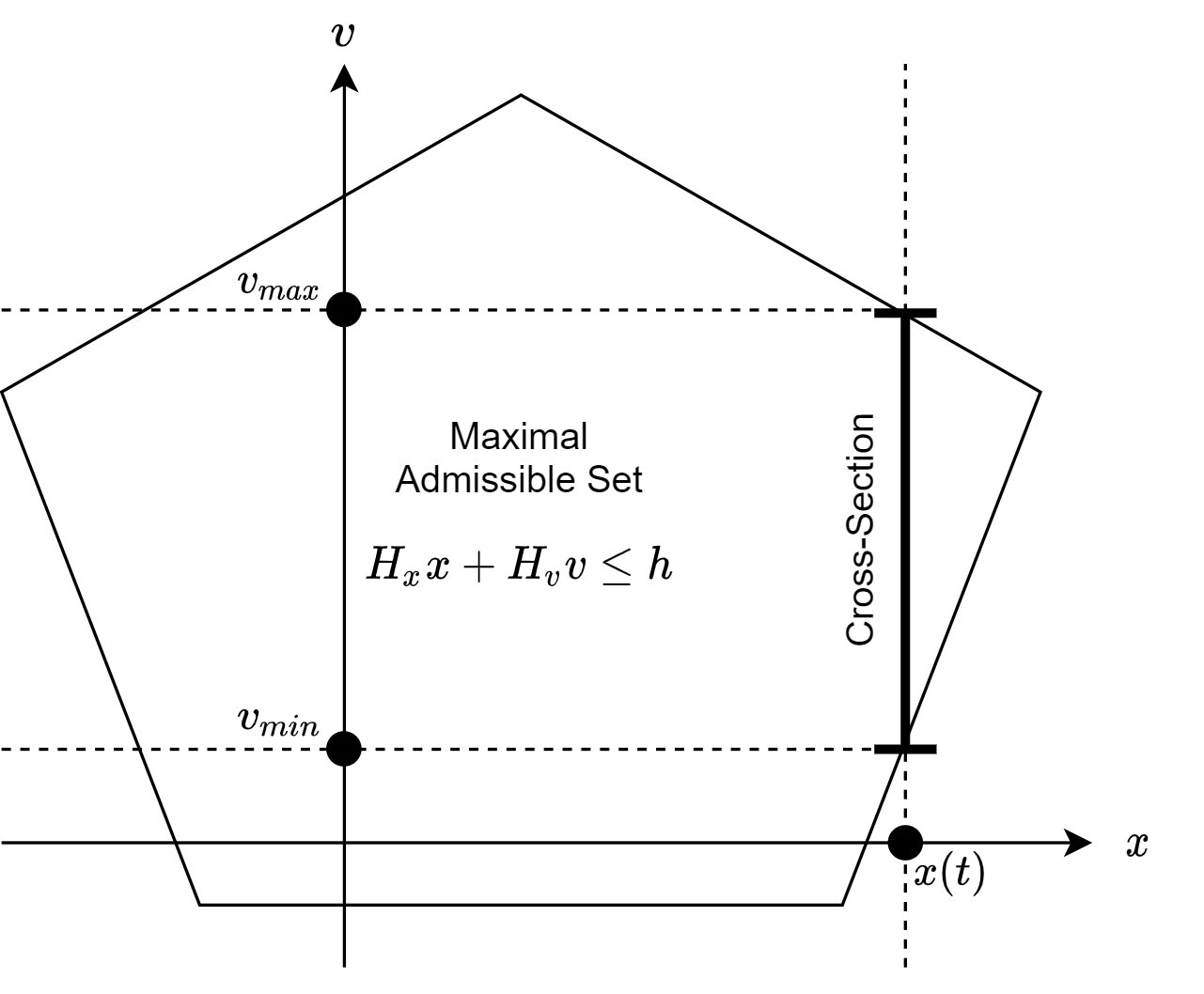}}
\caption{A graphical depiction of the Cross-Section RG.}
\label{fig:RGap-MASCS}
\end{figure}
In (\ref{eqn:RGap-CS-v}), if no $H_{v,j}$ satisfies $H_{v,j} > 0$ ($H_{v,j} < 0$), we assign a value of $M$ ($-M$) to $v_{max}[k]$ ($v_{min}[k]$), where $M$ is a very large number.  A graphical representation of {\eqref{eqn:RGap-CS-v}} 
is depicted in Fig.~\ref{fig:RGap-MASCS}, where the polygon represents $\widetilde{O}_\infty$. A cross-section of the MAS is defined for the known state, $x[k]$, which yields $v_{min}[k]$ and $v_{max}[k]$. However, to avoid exceedingly large values of $v$ and to ensure that the CS-RG output can be feasibly implemented by the hardware, $v[k]$ is further adjusted to $v_{lim}^{(-)}[k]$ and $v_{lim}^{(+)}[k]$, which are the lower limit and upper limit on the applied reference, respectively. These are defined to be functions of the desired reference, as follows:
\begin{equation}
    v_{lim}^{(\pm)}[k] = (1\pm\alpha)r[k]
    \label{eqn:RGap-CS-RG-bounds}
\end{equation}

\noindent where $\alpha$ is the fraction of the desired reference $r$ within which the CS-RG is allowed to increase (i.e., overshoot) or decrease (i.e., undershoot) the reference signal.

}

\begin{figure}
\vspace{-0.4cm}
\centerline{\includegraphics[width=1\linewidth]{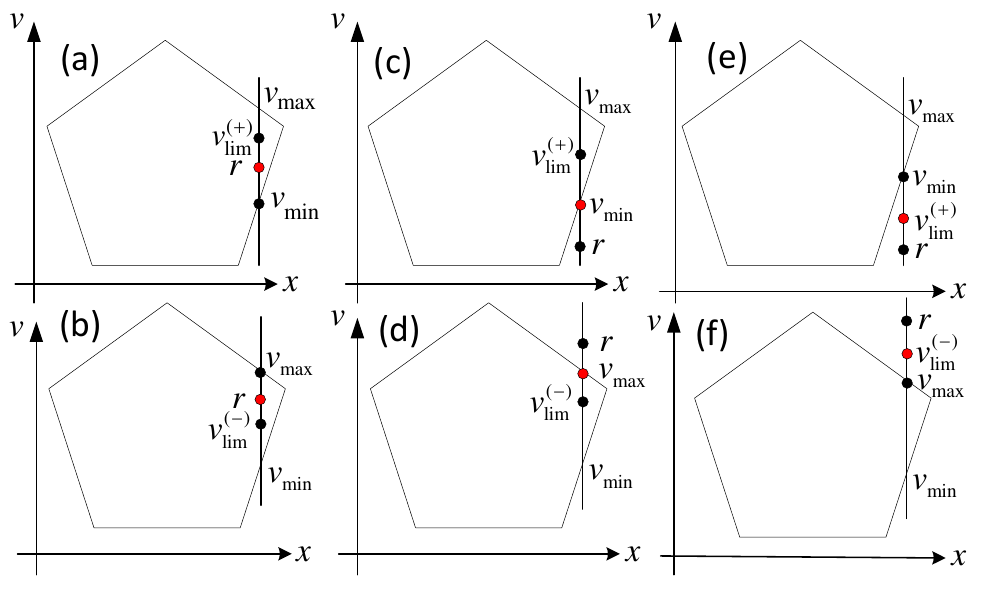}}
\caption{A graphical representation of the CS-RG equation presented in (\ref{eqn:CS-RG-algorithm}), where the red dot represents the applied reference by the CS-RG}
\label{fig:RGap-Formul}
\end{figure}

{In summary, the CS-RG algorithm works as follows. At each timestep $k$ of the control loop, the CS-RG first computes $v_{min}[k]$ and $v_{max}[k]$ using \eqref{eqn:RGap-CS-v}. The CS-RG then calculates $v[k]$ as follows:
\begin{equation}
v\left[k\right]={\mathrm{min} \left({\mathrm{max} \left(r[k],v_{t,min}[k]\right)\ },v_{t,max}[k]\right)\ } 
    \label{eqn:CS-RG-algorithm}
\end{equation}
where
\begin{equation*}
\begin{split}
v_{t,min}[k]={\mathrm{min} \left(v_{min}[k],v^{\left(+\right)}_{lim}[k]\right)\ }\\ 
v_{t,max}[k]=\mathrm{max}\mathrm{}\left(v_{max}[k],v^{(-)}_{lim}[k]\right)
\end{split}
\end{equation*}
If the desired reference $r[k]$ is already within the interval $[v_{min}[k], v_{max}[k]]$, then no constraint violation is predicted and the applied reference is $v[k] = r[k]$ (shown in Figs. \ref{fig:RGap-Formul}a and b). If $r[k]$ is outside this range, then it is adjusted to the constraint-admissible limits $v_{min}[k]$ and $v_{max}[k]$, provided that these limits are not excessively large or small and can be feasibly implemented with the available hardware (shown in Figs. \ref{fig:RGap-Formul}c and d). Alternatively, if that is not the case, i.e., $v_{min}[k]$  ($v_{max}[k]$) is excessively large (small), $r[k]$ is adjusted to $v^{\left(+\right)}_{lim}[k]$ ($v^{\left(-\right)}_{lim}[k]$), which is chosen to be within the hardware limitations (shown in Figs. \ref{fig:RGap-Formul}e and f). It is important to note that in this scenario, $v^{\left(\pm\right)}_{lim}[k]$ does not satisfy the constraint. However, in a CC-RG strategy, which is introduced in the next section, this adjustment aids in minimizing modifications to the references following the CS-RG operation.

}

Given the discontinuity in the CS-RG formulation, introducing noise into $r[k]$ may lead to the CS-RG toggling between $r[k]$ and $v^{\left(+\right)}_{lim}[k]$ or $v^{\left(-\right)}_{lim}[k]$, such as when the system state moves along the constraint boundary with a small noise on the input signal. To mitigate these effects, the partitioning of the MAS in~\eqref{eqn:RGap-CS-v} into $H_{v,j} > 0$ and $H_{v,j} < 0$ should involve a small tolerance to replace discontinuous toggling:
$H_{v,j} \geq \text{tol}$ and $H_{v,j} < -\text{tol}$. If $H_{v,j}=0$, then the constraint for that time-step is not dependent on $v[k]$ at all, being always satisfied or always violated depending on $x[k]$, so it has no {importance} on calculating $v[k]$ and can be skipped.

{

Before implementing the CS-RG on a closed-loop system, it must first be determined if the system will remain stable. If the system in \eqref{eqn:system_cont} is asymptotically stable, then that system with a CS-RG computing an applied reference signal based on \eqref{eqn:CS-RG-algorithm} will be BIBO stable due to the bounds on $v[k]$ between a finite maximum overshoot $v_{lim}^{(+)}[k]$ and undershoot $v_{lim}^{(-)}[k]$ of the desired reference set by the value of $\alpha$. Asymptotic stability for constant $r[k]$ is more difficult to prove because the applied reference is no longer monotonic, which is the case in standard RG. However, numerical simulations described later show that stability is not a concern.
}

\subsection{CC-RG Simulations}
\label{section:RGap-sim}

With the capability to govern multiple inputs in general directions, the novel CC-RG was then {applied to and} tested on the ninth-order nonlinear FCS model. { As we demonstrated that the FCS air-path has a low degree of nonlinearity in Section \ref{section:model-O9}, the assumption of a linear system required to prove BIBO stability of the Cascade RG and CS-RG (which together form the CC-RG) remain reasonable.}

In the case of the FCS OER constraint problem, stack current $I_{st}$ would be placed last in the Cascade RG so that stack current load is only modified after pressure and flow are modified as much as possible. The flow set-point RG uses a CS-RG. The pressure set-point RG uses the same algorithm as the {load governor} (\ref{eqn:kappa_for_RG}), as pressure set-points correlate negatively with OER. This is because higher pressure results in greater resistance to airflow into the cathode, and because steep pressure requests can cause transient ETB closings that briefly but severely restrict airflow. Thus, governing supply manifold pressure is still necessary to maintain the minimum OER constraint in transient.

The final output of the flow RG is bounded within $10\%$ {above} the desired reference $r$ { by setting $\alpha=0.1$ in \eqref{eqn:RGap-CS-RG-bounds}}, but this limit can be adjusted based on hardware and control specifications\footnote{
For the specific application of oxygen starvation protection for FCS, there is no maximum flow 
when strictly considering OER constraints. This may not remain true for surge/choke constraints as explored in \cite{vahidi_constraint_2005},\cite{vahidi_constraint_2007}, and unlimited flow is not a realistic way to operate physical compressors in any case.}.

The Cascade RG applied to the OER problem of an FCS relies on the assumption of monotonicity in the dynamics from the three references (i.e., $I_{st}$, $W_{cp}$, $p_{sm}$) to the OER constraint. Based on the physics of the FCS airpath, the following three assumptions can be made: (1) The higher the compressor mass flow, the further the system is from violating the OER constraint. (2) The higher the stack current and (3) the higher the cathode supply manifold pressure, the closer the {system} is to violating the OER constraint. This monotonicity allows each sub-RG in the cascade to reduce the action required of all the RG's ``downstream.'' { This property allows the Cascade RG to maintain the constraint in conditions where some of the sub-RGs could not enforce if operating on their own, such as slowing down the pressure set-point and current when the flow RG saturates at $10\%$ above the desired reference.}

\begin{figure}
\centerline{\includegraphics[width=0.9\linewidth]{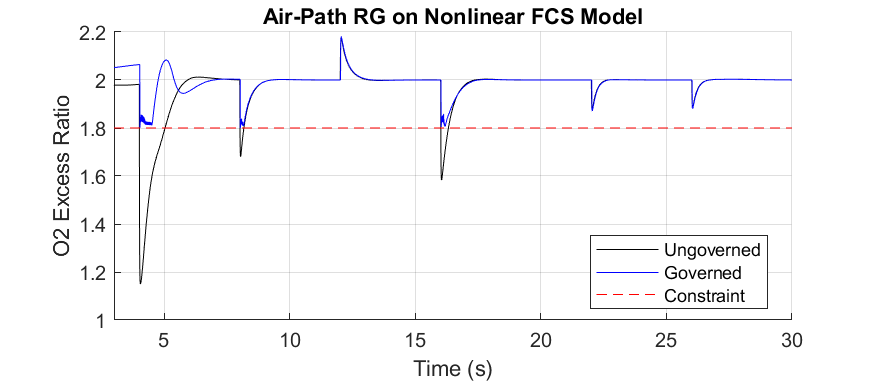}}
\caption{The OER responses to current steps, without and with the CC-RG on the ninth-order nonlinear model.}
\label{fig:RGap-OER}
\end{figure}

Fig.~\ref{fig:RGap-OER} shows the OER response without and with the CC-RG. As with the previous RG on stack current, the constraint is maintained at all times. Due to the presence of multiple coupled dynamics being governed independently, the transient OER response is not as smooth. To determine if this is due to an oscillation or instability, the next figure shall present the stack current input and output response of the FCS air-path.

\begin{figure}
\centerline{\includegraphics[width=1\linewidth]{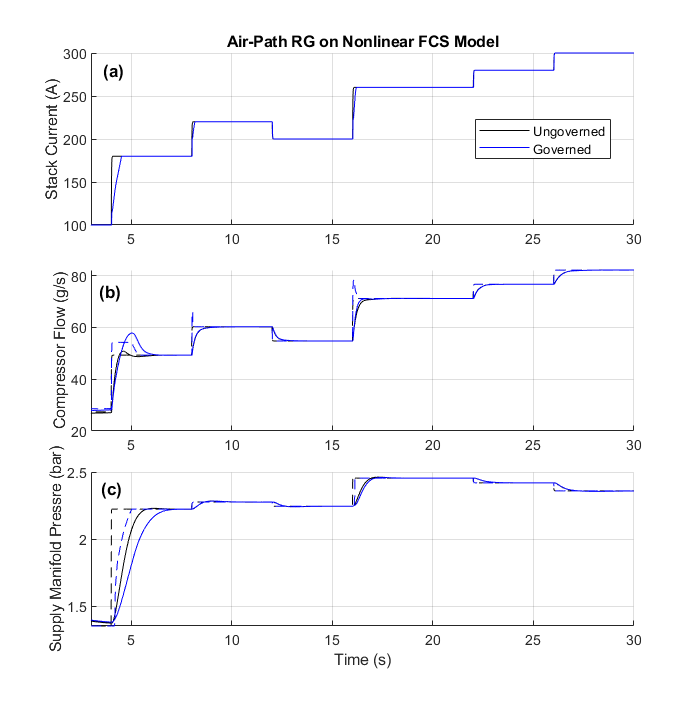}}
\caption{The stack current, compressor flow, and cathode inlet pressure, without and with the CC-RG on the ninth-order nonlinear model. For bottom two plots, dashed lines are set-points, solid lines are outputs.}
\label{fig:RGap-ry}
\end{figure}

Fig.~\ref{fig:RGap-ry} shows the resulting references and tracking performance of the FCS air-path without and with the CC-RG. Comparing Fig.~\ref{fig:RGap-ry}a to Fig.~\ref{fig:RG-OER} shows the CC-RG slows down the stack current far less than the {load governor}. Fig.~\ref{fig:RGap-ry}b shows how the Cross-Section RG speeds up the compressor mass flow set-point so that OER can be maintained at or above its lower limit, even if current is still being drawn relatively rapidly. Finally, Fig.~\ref{fig:RGap-ry}c shows that the cathode supply manifold pressure set-point is also modified to maintain the OER constraint. This is critical because when a steep rise in pressure is requested, the ETB may open more slowly or even close briefly and restrict the flow of oxygen. Each of these three reference signals and system outputs behave without oscillations or instability. Therefore, the ``ragged'' OER response in Fig.~\ref{fig:RGap-OER} is the result of the phase mismatch between states that influence OER. To further verify this, the actuator plots are also examined.

\begin{figure}
\centerline{\includegraphics[width=1\linewidth]{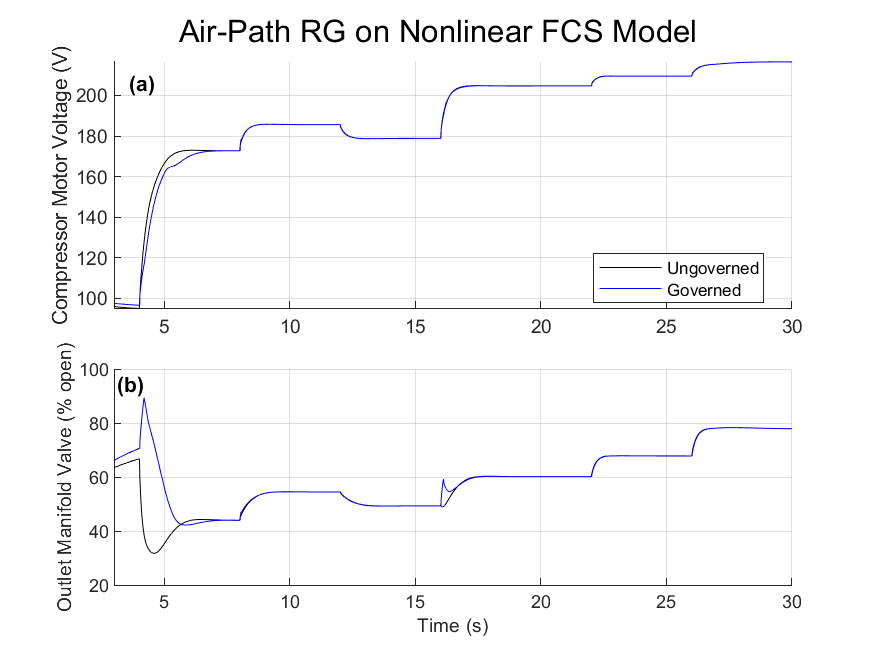}}
\caption{The compressor motor voltage and the ETB opening without and with the CC-RG  on the ninth-order nonlinear model.}
\label{fig:RGap-u}
\end{figure}

Fig.~\ref{fig:RGap-u} shows the actuator commands without and with the CC-RG. Despite the governed case requesting compressor mass flow more rapidly, the compressor action with the RG remains smooth and is only barely more aggressive than the ungoverned case, as shown in Fig.~\ref{fig:RGap-u}a. In Fig.~\ref{fig:RGap-u}b, the ETB in the governed case stays more open, or even opens rapidly, in transient in order to reduce the flow restriction when increased mass airflow is needed. An added benefit of using the ETB for brief increases in flow is that the compressor does not have to consume as much electricity to attain higher flow  rates, thus reducing parasitic losses.

In most cases, overshoot is not desirable for general tracking control. Using the CS-RG on the flow set-point represents an engineering trade-off to introduce calculated amounts of transient overshoot to prevent OER constraint violation. Thus, the maximum overshoot setting for the CS-RG is an added degree of freedom to trade-off flow tracking performance against OER constraint satisfaction. An alternative approach to using the CS-RG is to add overshoot via a dynamic feedforward map to generate flow set-points, which would augment or replace the static feedforward map (\ref{eqn:sFF-Wcp}). However, this approach lacks the constraint awareness of the RG, and thus may introduce flow overshoot when there is no risk of OER constraint violation. In contrast to an LTI dynamic feedforward map, the CS-RG introduces the minimum amount of overshoot that satisfies the constraints, including zero overshoot when the constraint will not be violated.

\begin{figure}
\centerline{\includegraphics[width=0.9\linewidth]{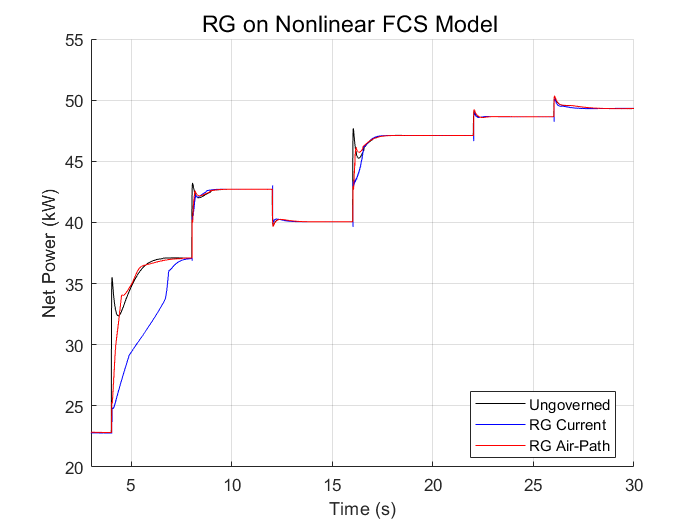}}
\caption{Net power response without RG, with the load governor, and with the CC-RG on the ninth-order nonlinear model.}
\label{fig:RG-both-Pnet}
\end{figure}

Fig.~\ref{fig:RG-both-Pnet} superimposes the net power response of the {load governor} and the novel CC-RG when tested on the same series of stack current steps. Of the two types of RG, the CC-RG more closely preserves the ungoverned net power response. This is because the CC-RG enables a faster current draw from the fuel cell stack.
Furthermore, the deliberate compressor flow overshoot for maintaining the OER constraint is primarily achieved with ETB actuation rather than compressor motor actuation, as shown in Fig.~\ref{fig:RGap-u}. By reducing flow restriction rather than spooling up the compressor, the CC-RG does not suffer from increased parasitic power losses to the compressor. This phenomenon further demonstrates the value of adding an ETB to the FCS air-path control problem in general.

{ The implementation of the CC-RG presented in this study assumes that compressor mass airflow and cathode inlet pressure are independently controllable. This was achieved by adding the cathode outlet manifold ETB to the FCS air-path control problem. However, it might be interesting to study the application of the CC-RG on an FCS with only the compressor mass airflow being tracked to a reference and no pressure set-point. This FCS architecture may be more common in practice. In this case, the CC-RG would have only two sub-RG's, a compressor mass airflow CS-RG followed by a stack current load governor. A potential drawback to this approach is that speeding up the flow would rely on spooling up the compressor rather than opening up the ETB (see Fig.~\ref{fig:RGap-u} at 4 seconds and 16 seconds for this phenomenon), thus increasing the compressor's parasitic load and possibly degrading net power rise time.}

The benefits of the CC-RG over the {load governor} are quantified at the end of the next section when both strategies are tested on realistic drive-cycles.

\section{Drive Cycle Simulations}

\label{section:drive-cycle}

\begin{figure}
\centerline{\includegraphics[width=1.0\linewidth]{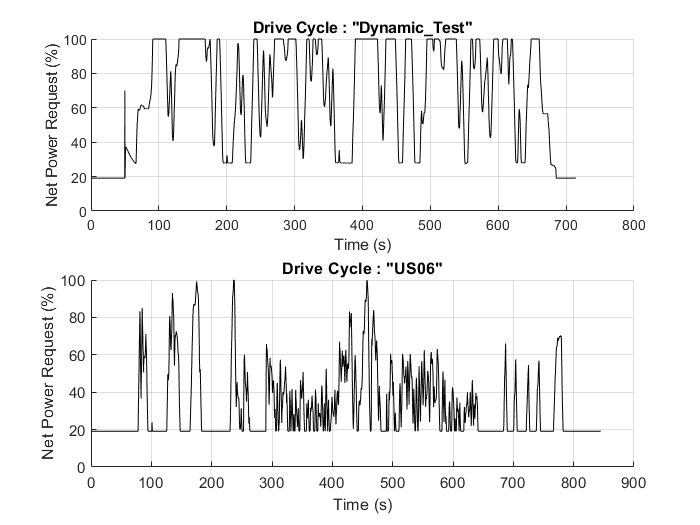}}
\caption{The net power requests for each drive-cycle.}
\label{fig:drvcyc-req}
\end{figure}

In order to quantify the performance of the proposed fuel cell control framework (shown in Fig.~\ref{fig:Intro-BlockHighLevel-MIMO}) in practical vehicle applications, the power requests from two real-world drive cycles are applied. The two drive cycles, referred to as ``Dynamic Test'' and ``US06'', are depicted in Fig.~\ref{fig:drvcyc-req}. In general, ``Dynamic Test'' spends more time at high power, while ``US06'' spends more time at low power. In the figure, power requests are represented as percentages rather than in units of power, due to the confidential nature of the data. Because these drive cycles were represented as power requests, a simple SISO power IMC was added to generate a stack current request given a net power request. The SISO power IMC was generated from a transfer-function plant model following the steps in Section \ref{section:IMC}, but in those cases $\Tilde{G}(s)$ would be a model of the entire closed-loop air path system, including the air-path controller.

\subsection{SISO vs MIMO Control}
\label{section:drive-cycle-control}

The MIMO IMC and the pressure set-point LUT were then tested on drive-cycle simulations. A SISO IMC controlling only the compressor motor voltage and tracking only the compressor mass flow (i.e., with no ETB actuation and no pressure tracking, designed with $\Tilde{G}(s) = G_{11}(s)$ from (\ref{eqn:IMC-plant-u-matrix})) was also developed and simulated on the same drive-cycles, as a point of comparison for the MIMO configuration.

Note that the primary source of efficiency improvements are not meant to come from the specific tuning nor the dynamics of the air-path controller. Rather, with the added ETB actuator and pressure tracking as new degrees of freedom, the FCS can be moved to an operating region where the compressor and the stack operate more efficiently than in the SISO case. This operating region is described via the pressure set-point map in Section \ref{section:set-point-maps}.

\begin{table}
\caption{Drive-Cycle Simulation Hydrogen Fuel Consumption}
\begin{center}
\begin{tabular}{|c|c|c|c|}
\hline
\textbf{Drive Cycle} & \textbf{SISO}& \textbf{MIMO}& \textbf{MIMO Percent Saved} \\
\hline
``Dynamic Test'' & 519.8 $gH_2$ & 481.5 $gH_2$ & 7.36 \%  \\
\hline
``US06'' & 253.8 $gH_2$ & 249.9 $gH_2$ & 0.81 \%  \\
\hline
\end{tabular}
\label{tab:drvcyc-H2}
\end{center}
\vspace{-0.6cm}
\end{table}

The resulting hydrogen fuel consumption from each simulation is shown in Table~\ref{tab:drvcyc-H2}. This is calculated as the total mass of dry hydrogen that has flowed out of the hydrogen supply. In both drive-cycles, the MIMO air-path configuration showed lower hydrogen fuel consumption than the SISO airpath configuration. These benefits are more pronounced for high-load scenarios, of which the ``Dynamic Test'' drive cycle is more representative than ``US06''.

\subsection{OER Constraint Management}
\label{section:drive-cycle-RG}

\begin{figure*}
\centerline{\includegraphics[width=0.8\textwidth]{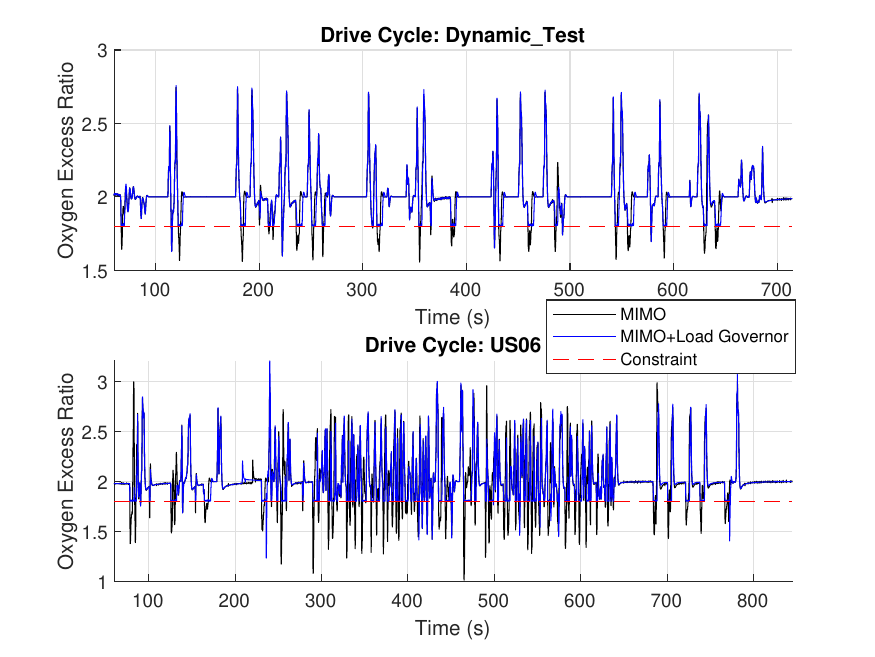}}
\caption{The OER responses of drive-cycle, without and with load governor on the ninth-order nonlinear model.}
\label{fig:drvcyc-OER}
\end{figure*}

\begin{figure*}
\centerline{\includegraphics[width=0.8\textwidth]{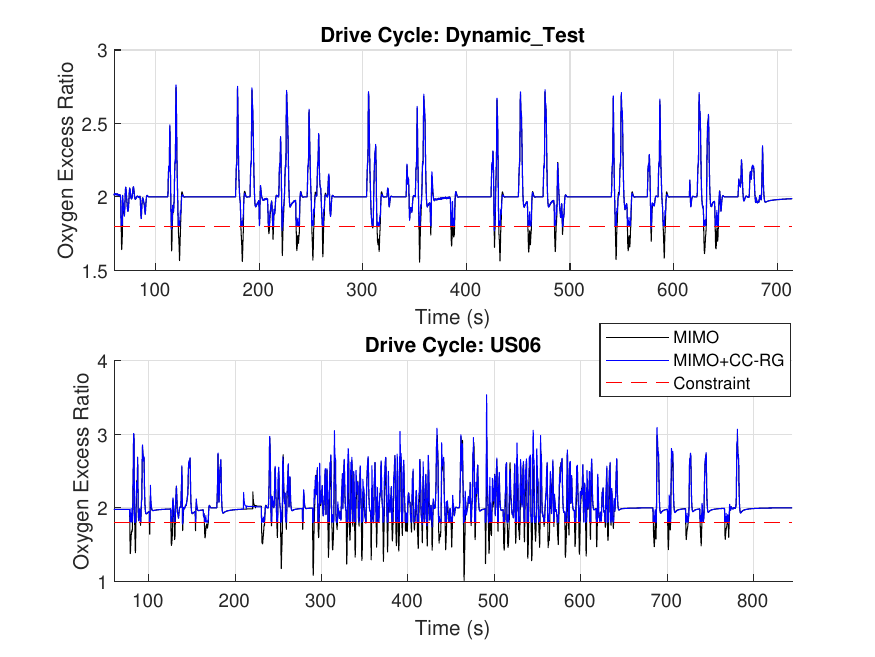}}
\caption{The OER responses of drive-cycle, without and with the CC-RG on the ninth-order nonlinear model.}
\label{fig:drvcyc-OER-RGap}
\end{figure*}

The MIMO air-path controller with both RG variations was also tested on both drive cycles. Fig.~\ref{fig:drvcyc-OER} shows the OER response of the MIMO system without and with the {load governor}. Due to the nonlinearity of the LUT mapping from the stack current to the pressure setpoint, the system with the RG may violate the OER constraint when a large pressure rise is requested, as this results in transient ETB closing. Fig.~\ref{fig:drvcyc-OER-RGap} shows the OER response of the MIMO system without and with the CC-RG. Because the CC-RG is placed after the set-point map, it does not have to approximate the nonlinear pressure LUT, and thus experiences less frequent and lower magnitude constraint violations. { By contrast, the traditional load governor does not always accurately predict OER constraint violation owing to its linear approximation of the nonlinear pressure LUT. Even if the minimum OER constraint can be maintained at all times with a sufficiently slow increase in stack current, the applied stack current computed by the load governor is limited by the accuracy of its prediction model.}

\begin{table}
\caption{Drive-Cycle Simulation Worst-Case OER}
\begin{center}
\begin{tabular}{|c|c|c|c|}
\hline
\textbf{Drive Cycle} & \textbf{Ungoverned} & {\textbf{Load Governor}} & {\textbf{CC-RG}}\\
\hline
``Dynamic Test'' & 1.56 & 1.60 & 1.77 \\
\hline
``US06'' & 1.01 & 1.23 & 1.74 \\
\hline
\end{tabular}
\label{tab:drvcyc-OER}
\end{center}
\end{table}

As shown in Table~\ref{tab:drvcyc-OER}, the presence of an RG reduces the magnitude of the worst-case transient OER in both drive cycles. The CC-RG showed the least worst-case OER constraint violation. The CC-RG experiences reduced modeling error because it is placed after the pressure set-point map, and therefore does not have to linearize the LUT. By artificially increasing the OER constraint to higher values, e.g., 1.9, these constraint violations can be alleviated. Even with plant-model mismatch and testing on a realistic drive cycle, the RG still demonstrates a significant improvement in transient OER.

\begin{table}
\caption{Drive-Cycle Simulation Net Power Mean Absolute Percent Error}
\begin{center}
\begin{tabular}{|c|c|c|c|}
\hline
\textbf{Drive Cycle} & \textbf{Ungoverned} & {\textbf{Load Governor}} & {\textbf{CC-RG}}\\
\hline
``Dynamic Test'' & 1.67 \% & 3.47 \% & 1.75 \% \\
\hline
``US06'' & 3.47 \% & 8.02 \% & 4.34 \% \\
\hline
\end{tabular}
\label{tab:drvcyc-Pnet}
\end{center}
\vspace{-0.4cm}
\end{table}

Table~\ref{tab:drvcyc-Pnet} compares the Mean Absolute Percent Error (MAPE) on net power of the ungoverned (MIMO) system against the {load governor} and the CC-RG. As a result of the CC-RG more closely preserving the desired stack current request, it demonstrates net power tracking error only slightly higher than the ungoverned case. By comparison, the presence of the {load governor} approximately doubles the average net power tracking error of the ungoverned case. Therefore, the novel CC-RG has the potential to improve the constraint awareness of the FCS controller while incurring less power tracking error than the {load governor}.

\begin{table}
\caption{Drive-Cycle Simulation Normalized Execution Time}
\begin{center}
\begin{tabular}{|c|c|c|c|}
\hline
\textbf{Drive Cycle} & \textbf{Ungoverned} & {\textbf{Load Governor}}& {\textbf{CC-RG}}\\
\hline
``Dynamic Test'' & 8.33 & 3.71 & 1.34  \\
\hline
``US06'' & 10.02 & 3.95 & 1.24  \\
\hline
\end{tabular}
\label{tab:drvcyc-time}
\end{center}
\end{table}

Table~\ref{tab:drvcyc-time} compares the normalized execution time of each drive-cycle simulation. A normalized time of 1 means equal to real-time, while greater than 1 means faster than real-time. These simulations were run on a Dell Laptop with 8 cores (10th Generation Intel Core i9-10885H processor) running at 2.40 GHz and 32GB of RAM. Matlab and Simulink software was used, and the simulations used a variable-step, variable-order solver (``ODE15s'') with a fixed RG sample time of 20ms. { As these simulations were obtained on a PC with a high-end processor, rather than an MCU or ECU, these results are less intended to determine the real-time speed of these approaches, but rather intended to compare the computational complexity of each approach to one another. If the CC-RG is not computationally efficient in real-time, one can remove ``almost redundant'' rows of the MAS or increase the sample time.} As can be seen, the MIMO IMC and both RG implementations ran faster than real-time. The ``slowest'' execution was with the novel CC-RG, which runs between 1 and 2 times as fast as real-time. This may be due to having to run three RG in series at each timestep (to support this possibility, the {load governor} runs about three times as fast). Determining and reducing the computational cost of the cascade RG and the CS-RG algorithms may be the subject of future work.

\section{High-Fidelity Simulations}

\label{section:highfidelity}

This section {tests} the MIMO LTI IMC and the novel CC-RG { on a proprietary high-fidelity FCS model at Ford Motor Company}. The primary findings are that the IMC and the CC-RG remain stable and effective even on a higher-fidelity nonlinear system.

\begin{figure}
\centerline{\includegraphics[width=0.9\linewidth]{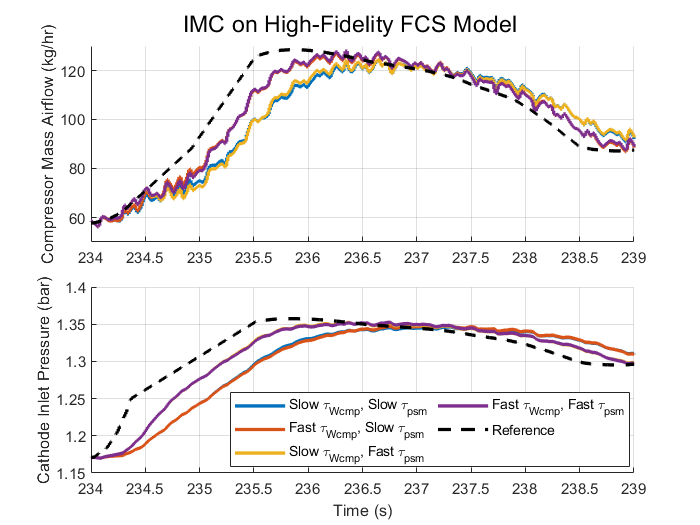}}
\caption{The compressor mass flow and the cathode inlet pressure responses on the high-fidelity model for four IMC tunings (set-point is black dashed line).}
\label{fig:high-fidelity-IMC-Wcp}
\end{figure}

\begin{figure}
    \centerline{\includegraphics[width=0.9\linewidth]{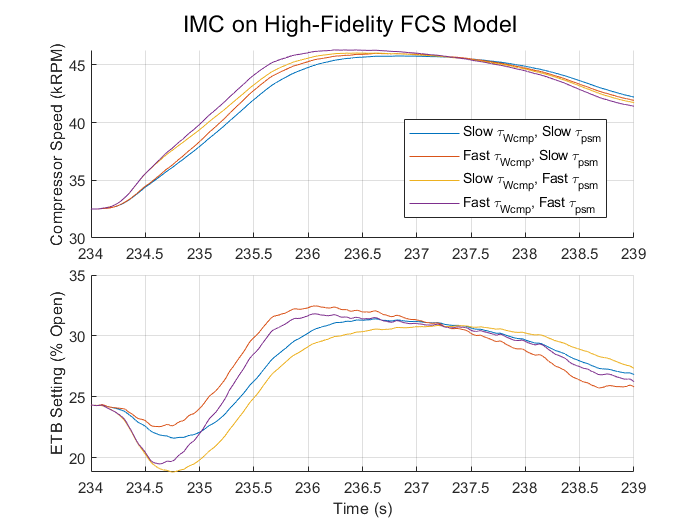}}
    \caption{The compressor speed and the ETB actuators on the high-fidelity model for four IMC tunings.}
    \label{fig:high-fidelity-IMC-u}
\end{figure}

Fig.~\ref{fig:high-fidelity-IMC-Wcp} shows the compressor mass flow and the cathode inlet pressure responses. The legend in this figure is omitted to hide the exact tuning parameters used for each controller, but the focus here is on the IMC itself and not on its tuning. The air-path set-points were generated with a built-in map based on the power request for a ``US06'' drive cycle. As this section will not discuss long-term performance of this simulation (e.g., the summary statistics tabulated in Section \ref{section:drive-cycle}), only this zoomed-in portion is shown. The IMC in this case is not identical to the controller presented in Section \ref{section:IMC}, but was built via system-identification on the high-fidelity model, followed by discretization to the Ford controller's sample time.

Furthermore, Fig.~\ref{fig:high-fidelity-IMC-u} shows stable actuator commands. Note that the Ford plant model accepts compressor shaft velocity rather than compressor motor voltage as a control input. Despite a noisier output response and the addition of more complex plant dynamics, the outputs show stable behavior and convergence towards the set-points, with a rise time in the 1 to 2 second range.

\begin{figure}
\centerline{\includegraphics[width=1.0\linewidth]{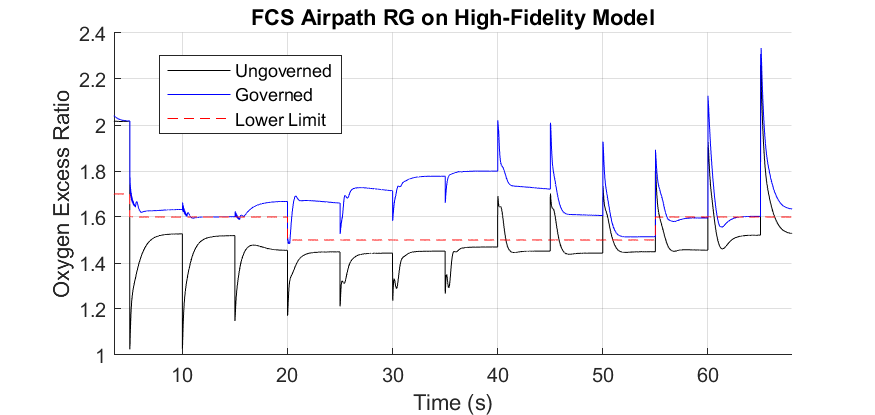}}
\caption{The OER response on the high-fidelity model without and with the CC-RG.}
\label{fig:high-fidelity-OER}
\end{figure}

{ The CC-RG was tested on the high-fidelity FCS model with a series of 50 A, 5-second stack current steps from 50 A to 400 A from 0 to 35 seconds, then 400 A back down to 50 A from 50 to 75 seconds}. Fig.~\ref{fig:high-fidelity-OER} shows the OER of the high-fidelity simulations subjected to stack current steps, without and with a cascade CC-RG.  { Unlike the open source model employed in previous sections, the} minimum OER constraint { of the proprietary Ford Model} varies as a function of stack current.

As can be seen, the governed case violates the minimum OER constraint far less frequently and less severely than in the ungoverned case. It is also worth noting that the compressor mass flow set-points (confidential) used with this model were not constraint-admissible at steady state, unlike the case with the ninth-order model discussed in previous sections of this paper. Due to greater plant-model mismatch, the RG exhibited some conservative behavior in the middle of the simulation, not hitting the OER constraint. Despite this, the CC-RG still resulted in a substantial improvement in worst-case transient OER. In general, the high-fidelity model demonstrated the effectiveness of the IMC for fast reference-tracking and the CC-RG for OER constraint management.

\section{Conclusions and Future Work}
\label{section:conclusions}

This paper presents a PEMFC air-path control scheme that is fuel-efficient, high-performance, and constraint-aware. The LTI MIMO IMC exhibited fast and robust tracking of the compressor mass flow and cathode supply manifold pressure. Using the new pressure set-point map, the system ran at higher net efficiencies, reducing hydrogen fuel consumption.
Finally, the OER has been maintained above a desired minimum value using the novel multi-reference CC-RG that enables faster power delivery than a traditional load governor on stack current. Overall, the results of this paper can provide new tools to improve PEM-FCS efficiency and reliability through mutlivariable control techniques.

From here, future work can be done to refine or improve the methodology of the set-point generation to further increase system net efficiency. For example, future maps could incorporate subsystems beyond the air-path, or attempt to create a dynamic rather than static mapping. Further work can be done to study and improve the cascade RG and the CS-RG, such as formalizing their theoretical and practical limitations and quantifying and improving their computational efficiency. Finally, and perhaps crucially, future study can cover the interaction between the air-path subsystem and the other FCS subsystems (e.g., thermal, humidity, hydrogen supply, energy management). Such efforts would support efforts for yet more holistic control of the FCS.

\bibliographystyle{IEEEtran}
\bibliography{FC_Journal_Paper}

\end{document}